\documentclass[reqno,11pt]{article}

 \usepackage{epsfig}
 \usepackage{amsmath}
 \usepackage{amsfonts}
 \usepackage{amssymb}
 \usepackage{amsthm}

 \usepackage{setspace}
 \usepackage{fullpage}

\usepackage{epsfig}
\usepackage{wrapfig}

 \newcommand{\hide}[1]{}

\newcommand{\Xomit}[1]{}

 \def\qed{\ifmmode$\blacksquare$\else{\unskip\nobreak\hfil
 \penalty50\hskip1em\null\nobreak\hfil$\blacksquare$
 \parfillskip=0pt\finalhyphendemerits=0\endgraf}\fi\vspace{0.3cm}}

 \newtheorem{theorem}{Theorem}[section]
 \newtheorem{observation}{Observation}[section]
 
 \newtheorem{definition}{Definition}[section]
 \newtheorem{lemma}{Lemma}[section]
 \newtheorem{corollary}{Corollary}[section]

 \newcommand{\xif}{{\bf{\em{if~}}}}
 \newcommand{\xthen}{{\bf{\em{then~}}}}

 \newcommand{\xfi}{{\bf{\em{fi~}}}}

 \newcommand{\vone}{\vspace{.1in}}
 \newcommand{\vhalf}{\vspace{.05in}}

 \newcommand{\ceil}[1]{\left\lceil{#1}\right\rceil}

\begin{document}

\title{\Large Efficient Resource Oblivious Algorithms for Multicores
}
\author{Richard Cole ~\thanks{Computer Science Dept., Courant Institute
of Mathematical Sciences, NYU, New York, NY 10012.
Email: {\tt cole@cs.nyu.edu}.
This work was supported in part by NSF Grant CCF-0830516.}
\and Vijaya Ramachandran~\thanks
{Dept. of Computer Science, University of Texas, Austin, TX 78712. Email:
 {\tt vlr@cs.utexas.edu}. This work was supported in
 part by NSF Grant CCF-0830737.}
 }

 \maketitle

\pagenumbering{arabic}
\thispagestyle{empty}

\begin{abstract}
We consider
the design of efficient algorithms for
a multicore
computing environment with a global shared memory and $p$ cores,
each having a cache of size $M$, and with
data  organized in blocks of size $B$. We
characterize the class of
`Hierarchical Balanced Parallel (HBP)' multithreaded computations for
multicores.
 HBP computations are similar to the
hierarchical divide \& conquer algorithms considered in recent work,
but have some additional features that guarantee good performance even
when accounting for the cache misses due to false sharing.
Most of our HBP algorithms are derived from known cache-oblivious
algorithms with high parallelism, however we incorporate new techniques
that reduce the effect of false-sharing.

Our approach to addressing false sharing costs (or more generally,
block misses) is to ensure that any task that can be stolen shares
$O(1)$ blocks with other tasks. We use a gapping technique for
computations that have larger than $O(1)$ block sharing. We also
incorporate the property of limited access writes analyzed in
\cite{CR11},
and we bound the cost of accessing shared blocks on the execution
stacks of tasks.

We present the {\it Priority Work Stealing (PWS)} scheduler,
and we establish that, given a sufficiently `tall' cache,
PWS deterministically schedules several highly parallel HBP algorithms,
including those for scans, matrix computations
and FFT, with cache misses bounded by the sequential complexity, when
accounting for both traditional cache misses and
for false sharing.
We also present a list ranking algorithm with almost optimal bounds.
PWS schedules without using cache or block size information, and uses
knowledge of processors only to the extent of determining the available
locations from which tasks may be stolen; thus it schedules
resource-obliviously.

 \end{abstract}

\section{Introduction}

We consider the efficient scheduling of multithreaded
algorithms \cite{CLRS09}
in a multicore computing environment.
We model a multicore as consisting of $p$ cores (or processors)
with an arbitrarily large shared memory, where
each core has a private cache of size $M$. Data
is organized in blocks of size $B$, and the initial input of
size $n$ is in the main memory, in $n/B$ blocks.
Recently, there has been considerable work on developing efficient
algorithms for multicores
\cite{FS06,CR07,BC+08,CR08,CSBR09,BGS09,Va08,AGN08,DLS08}; many of
these algorithms are multithreaded.
\hide{
Some of these papers consider a hierarchical caching environment where
the first (lowest) cache level contains the private caches we consider
in this paper.
Other types of multicore algorithms are given in \cite{Va08,AGN08,L?08}.
}
An efficient multicore algorithm attempts to obtain both
 work-efficient  speed-up as well as cache-efficiency.
However, none of these prior results have addressed false sharing costs
when considering cache-efficiency.

\vhalf
\noindent
{\bf Cache Misses, False Sharing, and Block Misses.}
When a core $C$ needs a
data item $x$ that is not in its private cache, it reads in the block
$\beta$ that contains $x$ from main memory at the cost of one cache miss.
This new block replaces an existing block in the private cache, which
is evicted using an optimal cache replacement policy (LRU suffices for
our algorithms).
If another core $C'$ modifies an entry in
block $\beta$, then
with {\it cache coherence},
$\beta$ is invalidated in $C$'s cache, and the next time core $C$
needs to access data in $\beta$, an updated copy of
$\beta$ is brought into $C$'s cache.
In the absence of cache coherence, some method of assigning ownership
to a block is needed so that the updates to items in a block are
correctly performed. For concreteness we will assume the above
cache coherence protocol.

The delay caused by different cores
writing into the same block can be quite significant, and this is a
caching delay that is present only in the parallel context.
In particular,
consider a parallel execution in which two or more cores between
them perform multiple
accesses to a block $\beta$, which include
$x \geq 1$ writes.
These accesses could cause
$\Omega (b \cdot x)$ delay at every core accessing $\beta$, where
$b$ is the delay due to a single cache miss.
These costs might arise if two cores are
sharing a block (which occurs for example if data partitioning does not
match block boundaries) or
if many cores access a single block
(which could occur if the cores are all executing very small tasks).
Further, $x$ can be arbitrarily large unless
care is taken in the algorithm design.
We refer to any access
of a block that is not in cache due to the block being
shared by multiple cores as a {\it block miss}.

{\it False-sharing} is a common example of a block being shared across
cores with the result that block misses occur.
This term is usually applied to the
case when two cores write into different segments of an array where the
two segments share a block. In this case, as mentioned above, each of
the two cores may incur $\Theta(B)$ cache misses when writing their
portion of this block due to the block `ping-ponging' between the
two cores. We use the term {\it block miss} in this paper
as a more general term to include all types of delays due to block-sharing.

\vhalf
\noindent
{\bf Schedulers and Resource Obliviousness.}
We consider multithreaded algorithms that expose parallelism by
forking (or spawning) tasks, 
but make no mention of the
processor (or core) that needs to execute any given task, and do not
tailor the size of the task being executed to the cache size or
block size.
Such an algorithm is called a {\it multicore-oblivious}
algorithm in \cite{CSBR09}. The multicore algorithms in
\cite{CR07,BC+08,CR08,CSBR09} are all multicore-oblivious; however,
these papers design run-time schedulers which use their knowledge
of cache and task
sizes in order to obtain efficient multicore performance.

In this paper, we use the term {\it resource-obliviousness} to refer to
an execution of a multicore-oblivious algorithm by a scheduler that
{\it does not} use knowledge of cache sizes or other parameters of the
multicore in its execution.
Earlier examples of resource-oblivious algorithms are in \cite{FS06,BGS09},
which use RWS (the
randomized work-stealing scheduler), but
these do not address the  cost of false sharing.
However, it appears that
if we want resource oblivious execution, we must consider the effect of
block misses, since we cannot avoid steals of tasks of size smaller than
$B$ unless the scheduler knows the block size. But if such small tasks
are stolen when writes occur within such tasks, block misses can be
expected to occur. The bounds obtained in \cite{FS06,BGS09} for RWS are weaker
than those we obtain for our scheduler, PWS, even if we ignore
block misses. In a companion paper \cite{CR11} we analyze the performance
of RWS considering both cache and block misses.

\vhalf
\noindent
{\bf Our Contributions.}
 The contributions of this paper are two-fold.

1. First, we set up a framework to analyze -- and optimize for --
the caching overhead of both cache misses and
block misses
while also allowing for
high parallelism.

We identify a
basic primitive, the {\it balanced parallel (BP)} computation.
The BP computation is the basic building
block in {\it Hierarchical
Balanced Parallel (HBP)} computations, which are obtained through
sequencing and parallel recursion.
HBP computations have the {\it limited access} property for writes,
which requires
that any variable is written at most a
constant number of times.
We present techniques for reducing the cost of block misses, such as
{\it $O(1)$-block sharing}, and {\it gapping}.
We also use a result presented in
our
companion paper \cite{CR11} which bounds the number of access to any given
block in a class of algorithms with limited access writes that includes
HBP computations.

HBP computations are similar to the Hierarchical Divide and Conquer
(HD\&C) class in \cite{BC+08}. But they have
important differences, notably the limited access
requirement. We analyze HBP algorithms for scans, matrix transposition
(MT), Strassen's matrix multiplication, converting a matrix between row major
(RM) and bit interleaved (BI) layouts, FFT, list ranking and graph
connected components. Most of these are known algorithms,
but some are new and others are modified to conform to HBP and to
achieve low block miss cost.

On the algorithm analysis side, our
analyses are in terms of certain structural parameters of the algorithms.
On an input of length $n$,
these include the work $W(n)$, the depth or critical path length $T_{\infty}$,
the cache complexity in a sequential execution $Q(n,M,B)$, and new parameters:
$f(r)$, the cache-friendliness, and $L(r)$, a block sharing measure.
It suffices to determine these parameters in order to analyze the
algorithms.
Further, the algorithm design problem
can focus on minimizing these parameters\footnote{
We note that the recent sorting algorithm in \cite{CR10b} uses
the scheduling bounds developed here in its analysis.}.

\vhalf
2. Our second contribution is a new deterministic work-stealing scheduler,
the {\it Priority Work-Stealing Scheduler (PWS)},
which is tailored to perform well on HBP computations.
It achieves a lower caching overhead
due to steals than the bounds derived for RWS
in \cite{FS06,BGS09} for the case when block misses are
not considered, and in our companion paper \cite{CR11} when both
cache and block misses are considered.
For most of the algorithms we consider, we obtain
{\it optimal} cache miss overhead with a sufficiently tall cache,
considering both cache and block misses,
when the input is larger than the
combined sizes of the caches.
PWS is also a deterministic scheduler, for which we give a reasonably
simple distributed implementation.

\section{Computation Model}
\label{sec:comp-model}

We consider a class of multithreaded
parallel computations that
expose their parallelism  through binary forking
of  parallel tasks
(see, e.g.,~\cite{CLRS09}, Chapter 27).
Parallel tasks are scheduled on cores using a {\it work stealing} scheduler.

The basic unit
in our formulation
is a computation tree $T$ with
binary forking
of tasks, which forms the {\it downpass} of the computation.
The downpass is followed by an {\it up-pass} on a reverse tree where two
forked tasks join, and once the execution of the two forked
tasks is completed, the computation is continued by the task that
forked them.
A simple example, M-Sum, is shown below,
which computes the sum of the $n$ elements in
array $A$.

\begin{figure}[htbp]
\small{
M-Sum$(A[1..n], s)$ \hspace{1in} \% Returns $s = \sum_{i=1}^n A[i]$\\
\xif $n=1$ \xthen {\bf return} $s := A[1]$ \xfi \\
{\bf fork}(M-Sum$(A[1..n/2],s_1)$; M-Sum$(A[\frac{n}{2}+1 .. n],s_2))$\\
{\bf return} $s = s_1 + s_2$
} 

\end{figure}

Initially the root task for M-Sum's computation is given to a single core.
This root task corresponds to the entire computation of M-Sum on array
$A[1..n]$.
In a sequential execution, this computation
proceeds by ignoring fork commands. In a work stealing
multicore execution, subtasks
are acquired by other cores via task stealing.
To this end, each core $C$ has a task queue.
It adds forked tasks to the bottom of the queue, while tasks are stolen from the top of the queue.
So in particular, when  $C$, on executing a task
$\tau$, generates forked tasks $\tau_1$ and $\tau_2$,
it places $\tau_2$ on its queue, and continues with the execution of $\tau_1$.
When $C$ completes $\tau_1$, if $\tau_2$ is still on its queue, it resumes the execution of $\tau_2$,
and otherwise there is nothing on its queue so it seeks to steal a new task.
The core executing the last of $\tau_1$ and $\tau_2$ to finish
will complete the execution of $\tau$
in the up-pass.
While $C$ is executing task $\tau$ other cores will be acquiring work
by stealing tasks
from $C$'s task queue, and then in turn
will be generating their own task queues
from which further subtasks can be stolen.
We will refer to the portion of task $\tau$ executed by $C$ as the
{\it kernel} of task $\tau$.

We have outlined
the mechanism of work-stealing for a
simple tree-structured computation. However, work-stealing applies
to general DAG-structured computations,
and {\it randomized} work stealing
has been analyzed for DAGs represented by series-parallel graphs and
more general structures (e.g., \cite{BL99,ABB02}).
In this paper,
we consider algorithms whose computation DAG represents a
{\it Hierarchical Balanced Parallel (HBP)} computation, which is
defined in Section \ref{sec:alg},
and we establish that they perform very well when executed under
the {\it Priority Work Stealing (PWS)}
scheduler, which we introduce in Section \ref{sec:pws}.

\subsection{Cache Misses}
\label{sec:c-miss}

Work stealing causes execution of a multithreaded algorithm to
incur additional cache misses
over those incurred in a sequential execution, and it also
introduces block misses.
We introduce two parameters to express these costs,
the {\it cache-friendliness} function $f(r)$ and the
{\it block-sharing} function $L(r)$.
We introduce $f(r)$ here. We define $L(r)$ and discuss block misses
further in Section \ref{sec:b-miss}.

\begin{definition}\label{defn:friendly}
A collection of $r$ words of data is {\it $f$-cache friendly} if
they are contained in $O(r/B + f(r))$ blocks. An HBP computation is
$f$-cache friendly if for every task $\tau$
in the computation,
the sequence of words accessed by
$\tau$ is $f(|\tau|)$-cache friendly.
\end{definition}

For instance, $f(r)= 1$ if $\tau$ accesses an array stored in contiguous
locations; if $\tau$ access a 
$\sqrt r \times \sqrt r$ submatrix of a matrix stored in RM (i.e., row major),
then $f(r) = \sqrt r$.

Let $\tau$ be a task in a multithreaded
algorithm $\cal A$.
We will use the {\it size} of $\tau$ to
denote the number of words accessed by $\tau$.
If $\tau$ is stolen by a core $C$,
then $C$ incurs additional cache misses over the sequential execution,
since it will have to
read the possibly previously read data needed to execute $\tau$.
Define $Q_{\tau}$ to be the number of cache misses incurred by task $\tau$
in the sequential execution of $\cal A$.
The \emph{excess cache miss} caused by the steal of $\tau$
is defined to be the number of cache misses incurred by
$C$ in its execution of $\tau$
minus $c\cdot Q_{\tau}$, for a suitable constant $c\ge 1$.

A stolen task of size $M$ could have incurred no cache misses in an
execution in which
it was not stolen, but once the size of the stolen task reaches $2M$,
its execution when not stolen
would incur at least $M/B$ cache misses.
The following lemma makes this precise.

\begin{lemma}
\label{lem:c-miss-basic-bdd}
A stolen task $\tau$ incurs at most
$O(\min\{\frac{M}{B}, \frac{|\tau|}{B}\} +f(|\tau|)\})$
additional
cache misses compared to the
steal-free sequential computation.
If $f(|\tau|) = O(|\tau|/B)$
and $|\tau| \ge 2M$,
this is an
excess of 0 cache misses.
\end{lemma}
\begin{proof}
In the sequential execution of the algorithm, the execution of $\tau$ incurs
at least $Q_{\tau}=\max\{0,\frac{|\tau|}{B} - \frac{M}{B}\}$ cache misses
since $|\tau|$ data has to be accessed, of which at most $M$
is in cache.
$C$'s execution of $\tau$ incurs $O(\frac{|\tau|}{B} + f(|\tau|)) =
O(Q_{\tau} + \min\{\frac{|\tau|}{B}, \frac{M}{B}\} + f(|\tau|))$
cache misses.
For $|\tau| \ge 2M$, $Q_{\tau} \ge \frac{|\tau|}{2B}$, and if
$f(|\tau|) = O(|\tau|/B)$, then
$\min\{\frac{|\tau|}{B}, \frac{M}{B}\} + f(|\tau|)= O(Q_{\tau})$.
\end{proof}

\subsection{Block Misses}
\label{sec:b-miss}

We discuss here our basic set-up for coping with block misses.
As mentioned earlier, we assume that block misses are handled under
a cache coherence protocol whereby a write into a location in a shared
block $\beta$ by core $C$ invalidates the copy of $\beta$ in every
other cache that holds $\beta$ at the time of the write. This is done
so that data consistency is maintained within the elements of a block
across all copies in caches at all times.
There are other ways of dealing with block misses
(see, e.g., \cite{HP06}),
but we believe that the block miss cost 
with our invalidation rule
is likely as high as (or higher than) that
incurred by other mechanisms. Thus, our upper bounds should hold
for most of the coping mechanisms known for handling block misses.

A block miss occurs at a core $C$ when it has a block $\beta$ which
it shares with
one or more other cores, and it needs to read $\beta$ again because
another core wrote into a location in $\beta$, thereby invalidating
the copy of $\beta$ in $C$'s cache. The cost of such a block miss is
at least that of one cache miss, but it could be much larger, depending
on the number of cores that share $\beta$ and write into it;
in fact, the cost of a block miss could be unbounded in a scenario where
several cores repeatedly write into locations in the block,
if the system
mechanism for transferring access to the block does not ensure fairness.
Our analysis for bounding the cost of block misses does not make any
assumptions about the mechanism used for transferring accesses to a shared
block under writes. Hence the bounds we obtain are truly worst-case.

\begin{definition}
\label{def:block-delay}
Suppose that block $\beta$ is moved $m$ times
from one cache to another \emph{(}due to cache or block misses\emph{)}
during a time interval $T = [t_1,t_2]$.
Then $m$ is defined to be the\emph{ block delay} incurred by $\beta$ during $T$.

The \emph{block wait cost} incurred by a task $\tau$ on a block $\beta$
is the delay incurred during the execution of $\tau$
due to block misses when accessing $\beta$, measured
in units of cache misses.
\end{definition}

Note that the
block wait cost incurred by a task $\tau$ on a block $\beta$
is the delay incurred as measured in units of cache misses.
Clearly, the block delay of a block $\beta$ during a time interval $T$ is
an upper bound on the block wait cost incurred by any task on block $\beta$
during $T$.

We now define $L$-block sharing.

\begin{definition}\label{defn:block-sharing}
A task $\tau$
of size $r$
is \emph{$L$-block sharing},
if there are $O(L(r))$ blocks which
$\tau$ can share with all other tasks that
could be scheduled in parallel with $\tau$
and could access a location in the block
(these other tasks do not include subtasks of $\tau$).
A computation
has block sharing function $L$ if every task in it is $L$-block sharing.
\end{definition}

The following definition is from \cite{CR11}
\begin{definition}
\label{def:limited-use-var}
\cite{CR11}
An algorithm is \emph{limited-access} if each of its writable
variables is accessed $O(1)$ times.
\end{definition}

The two main algorithmic techniques that we use to
reduce the cost of block misses are to enforce $O(1)$-block sharing
and the limited access.
In some of the algorithms, we also
use a {\it gapping} technique to reduce
the block miss cost.
We will also
assume the following system property.
Whenever a core requests space
it is allocated in block sized units; naturally, the
allocations to different cores are disjoint and entail no block sharing.

\section{HBP Computations and Algorithms}
\label{sec:bp-comp-alg}

{\it Balanced parallel} (BP) computations, defined below, form
the backbone of our HBP algorithms.
Recall that
the {\it size} $|\tau|$ of
a task $\tau$ as the amount of data accessed by $\tau$.
Note that the size of a task is a positive integer.

It will also be helpful to specify the notions of local and global variables.
\begin{definition}
\label{def:glob-loc-var}
A variable $x$ declared in a procedure $P$ is called a \emph{local} variable
of $P$.
A variable $y$ accessed by $P$ and declared in a procedure $Q$ calling $P$
or used for the inputs or outputs of the
algorithm $\cal A$ containing $P$ is said to be \emph{global} with
respect to $P$.
However, note that $y$ would be a local variable of $Q$ if declared in $Q$.
\end{definition}

\begin{definition}\label{def:BP}
A  \emph{BP computation} $\pi$ is a limited access algorithm that is formed
from the downpass of a binary forking computation tree $T$
followed by its up-pass, and satisfies the following properties.
\begin{description}
\item[i.]
A task that is not a leaf performs only $O(1)$ computation before it forks its two children in
the downpass of the computation.
\item[ii.]
In the up-pass each task performs only $O(1)$ computation after the
completion of its forked
subtasks.
\item[iii.]
Each leaf node performs $O(1)$ computation.
\item[iv]
Each node declares at most $O(1)$ local variables.
\item[v.]
$\pi$ may also use size $O(|T|)$ global arrays for its input and output.

\item[vi.] \emph{Balance Condition.}
Let the height of $T$ is $h$;
let the root task, which is at level 0 in $T$, have size $r$;
let $\alpha$ be a constant less than 1; and let $c_1, ~c_2$ be constants with
$c_1 \leq 1 \leq c_2$. Then,
the size of any task $\tau$ at level $i$ in $T$ satisfies
$c_1 \cdot \alpha^i \cdot  r \leq |\tau| \leq c_2 \cdot \alpha^i \cdot r$.

\end{description}
\end{definition}

The {\it task head} of a task $\tau$ is the computation it performs
in part (i) in the above definition.

This definition of a BP computation requires sibling tasks to have essentially
the same size to within a constant factor.
However, a computation
in which these sizes are upper bounds on the actual size is sufficient
for our results, as long as this upper bound on the actual size is what
is used to compute the resource bounds.
Also, note that
any BP computation will have $\alpha \geq 1/2$;
all of our algorithms have $\alpha = 1/2$.

Later, in Section \ref{sec:pws-impl}, in order to reduce the
block wait costs in our scheduler implementation,
we will employ a variant of BP computations
which we call \emph{padded} BP computations.

\begin{definition}
\label{def:paddedBP}
A \emph{padded} BP computation is a BP computation in which each node $v$ in the
down-pass declares an array:
let $v$ corresponds to the start of a subtask $\tau$;
then $v$'s array is of size
$\sqrt{|\tau|}$.
\end{definition}
These arrays are present to ensure that the space used by successive nodes to
store their variables (other than the new array) are well separated;
this is what enables a reduction in block wait costs.

\vhalf
In Section
\ref{sec:pws-cache}, we
will use the following observation on the nature of stolen tasks
in a BP computation under work-stealing.
(This observation holds more generally for series-parallel computation dags.)

\begin{observation}
\label{obs:stolen-tasks}
Let $D$ be the computation dag for a
BP computation
$\Pi$ executing at a core $C$ under work stealing.
Let $v$ be the node in $D$ corresponding to the last task $\tau_v$
that was stolen from $C$ while it was executing $\Pi$, and let $P$
be the path in $D$ from the
root of $D$ to the parent of $v$.
Then, the set of tasks stolen from $C$ during its execution of $\Pi$
consists of some or all of the tasks corresponding to those nodes of $D$ that
are the right child of a node in $P$
but are not themselves on $P$.
Further, they are stolen in top-down order with respect to the path
$P$.
\end{observation}

\subsection{HBP Computations}\label{sec:hbp}

We now define the class of
HBP Computations.

\begin{definition}
\label{def:rec-alg}
A \emph{Hierarchical Balanced Parallel Computations (HBP)} is a limited
access algorithm that is one
of the following:
\begin{enumerate}
\item
 A Type 0 Algorithm, a sequential computation of constant size.

\item
A Type 1, or BP computation.
\item
A Type $i+1$ HBP, for $i \geq 1$.
An algorithm is a Type $i+1$ HBP if,
on an input of size $n$, it calls, in succession,
a sequence of $c \ge 1$ collections of $v(n) \ge 1$ parallel recursive
subproblems,
where each subproblem has size $s(n) \le n/b(n)$, with $b(n) > 1$;
further, each of these collections
can be preceded and/or followed by calls to HBP algorithms of type at most $t$.

Data is transferred to and from the recursive subproblems by means of
variables (arrays) declared at
the start of the calling procedure.
\item
A Type $\max\{t_1, t_2\}$ HBP computation results if it
is a sequence of two HBP algorithms of types $t_1$ and $t_2$.
\end{enumerate}

A \emph{Padded HBP} computation is an HBP computation in which each BP subcomputation is padded.
\end{definition}

\begin{definition}\label{def:HBP}
An HBP computation of type $t>1$ is
\emph{balanced} if 
the recursive problems at each level of recursion
all have sizes within a constant
factor of each other.
\end{definition}

For convenience, we will assume this constant factor in balanced
HBP computations to be $c_2/c_1$, where $c_1$ and $c_2$ are the constants
in Definition \ref{def:BP}.

All HBP algorithms we consider here are balanced, but the sorting algorithm
SPMS in our recent paper \cite{CR10b} is unbalanced.

The HBP class
is closely related to the {\it Hierarchical Divide and Conquer (HD\&C)}
 class in \cite{BC+08} (after the parallelism is exposed in the
HD\&C algorithms).
The HD\&C class was used in \cite{BC+08}
for a 3-level cache hierarchy with
a special scheduler that is {\it not} oblivious to cache parameters.
The main differences
between HBP and HD\&C
are that we allow sequencing of HBP computations
even at the top level, and we do not restrict the number of
subproblems that are called recursively to be bounded by
a constant; on the other hand we restrict the computation to be
limited access.

\vhalf
\noindent
{\bf Forking recursive tasks.}
The recursive forking of $v(n)$ parallel tasks in an HBP computation is
incorporated into the binary forking in our multithreaded set-up by
a BP-like tree of depth $\log_2 v(n)$. All nodes at
a given level have the same number of recursive subproblems, to within
a constant factor. Each leaf of this tree is a recursive subproblem.
By its construction such a BP-like tree will have $\alpha = 1/2$ in a
balanced HBP.

\subsection{HBP Algorithms}\label{sec:alg}

Our results in for PWS in Section \ref{sec:pws} establish that
$L(r)=O(1)$ is desirable, while
$f(r)=O(\sqrt r)$ suffices if have a standard tall cache $M \geq B^2$.
 Table 1 lists the HBP algorithms that we present and analyze in
this paper.
Most of these algorithms are adapted from known HD\&C
algorithms. All of them are limited access and have $f(r) = O(\sqrt r)$;
for Depth-n-MM, the original
algorithm in \cite{FLPR99} is not limited access, but it is converted
to being limited access by using local arrays for copying in \cite{CR11}.
Many of these algorithms also inherently have
$L(r) = O(1)$ (e.g.,
Scans, MT (Matrix Transposition) and Strassen (Matrix Multiplication)),
while others  are
modified through the gapping technique to reduce the block miss cost.

\begin{table}[t]
\label{table2}
\begin{center}
\begin{tabular}{|c||c|c|c||c|c|c|}
\hline
 {\footnotesize {\bf Algorithm}} &
 {\footnotesize {\bf Type }} &
{\footnotesize {\bf $f(r)$}} &
  {\footnotesize {\bf $L(r)$}} &
{\footnotesize {\bf $W(n)$ }} &
  {\footnotesize {\bf $T_{\infty}$ } } &
  {\footnotesize {\bf $Q(n,M,B)$ }}
\\\hline
\hline
Scans (MA, PS) & 1 & 1 & 1  & $O(n)$ & $O(\log n)$  & $O(n/B)$ \\ \hline
MT  & 1 & 1 & 1 & $O(n^2)$ & $O(\log n)$  & $O(n/B)$ \\ \hline
 Strassen   & 2 & 1 & 1  & $O(n^{\lambda})$ & $O(\log^2 n)$ & $n^{\lambda}/(B\cdot  M^{\frac{\lambda}{2} -1})$ \\ \hline
RM to BI & 1 & $\sqrt r$ & 1 &  $O(n^2)$ & $O(\log n)$  & $O(n^2/B)$ \\ \hline
Direct BI to RM & 1 & $\sqrt r$ & $\sqrt r$ & $O(n^2)$ & $O(\log n)$  & $O(n^2/B)$ \\ \hline
BI-RM (gap RM)  & 1 & $\sqrt r$ & gap & $O(n^2)$ & $O(\log n)$  & $O(n^2/B)$ \\ \hline
%
BI-RM for FFT & 2 & $\sqrt r$ & 1 & $O(n^2\log \log n)$ & $O(\log n )$  & $O(\frac{n^2}{B} \log_M n)$ \\ \hline
FFT & 2 & $\sqrt r$ & 1 &  $O(n \log n)$ & $O(\log n \cdot \log \log n)$  & $O(\frac{n}{B} \log_M n)$ \\ \hline
LR & 3 & $\sqrt r$ & gap &  $O(n \log n)$ & $O(\log^2 n \cdot \log\log n)$  & $O(\frac{n}{B} \log_M n)$  \\ \hline
CC & 4 & $\sqrt r$ & gap &  $O(n \log^2 n)$ & $O(\log^3 n \cdot \log^{(2)} n )$  & $O(\frac{n}{B} \log_M n \cdot \log n)$ \\ \hline
\hline
Depth-n-MM \cite{CR11} & 2 & 1 & 1 & $O(n^3)$ & $O(n)$ & $n^3/(B\sqrt M)$ \\ \hline
Sort \cite{CR10b} & 2 & $\sqrt r$ & 1  & $O(n \log n)$ & $O(\log n \cdot \log \log n)$  & $O(\frac{n}{B} \log_M n)$ \\ \hline
\end{tabular}
\end{center}
\caption{Basic parameters of the HBP algorithms we analyze.
Type refers to the HBP type, $f(r)$ is the cache-friendliness function,
and $L(r)$ is the block-sharing function. The bounds with $f(r)=\sqrt{r}$
assume a tall cache.
The input size is $n$, except for matrix computations, where the input
size is $n^2$.
For completeness, we include the known bounds for
work ($W(n)$),
critical pathlength ($T_{\infty}$), and sequential cache complexity
($Q(n)$).
} 

\end{table}

\vhalf
{\bf Scans},  including {\sc M-Sum} seen earlier, and {\bf MA} (Matrix
Addition) \cite{BC+08} can be implemented
as a single
BP computation.
{\it Prefix sums} ({\bf PS})
can be implemented as a sequence of two BP computations, where
the first BP computation
computes sums of disjoint subarrays of size $2^i$, for $i<\log n$,
and the
second BP computation computes the final output.
These are type 1 HBP computations with $f(r)=O(1)$, $L(r)=O(1)$.

\vhalf
{\bf Matrix Computations.}
For matrix computations, we assume that the matrix is in
the {\it bit interleaved (BI) layout}, which recursively
places the elements in the top-left quadrant, followed by
recursively placing
the top-right,
bottom-left, and bottom-right quadrants.
The advantage of the BI layout is that it results in BP tasks that
are $O(1)$-friendly,  and
have $O(1)$-block sharing,
which allows us to obtain good cache and block miss bounds.
We describe several methods to convert between the standard
{\it row major (RM)} layout and BI; these methods can be used in conjunction
with our algorithms for BI if the
input and output matrices are to be in RM.

\vhalf
{\bf MT} is matrix transposition when the $n\times n$ matrix is given in
the BI layout. When we expose the parallelism in
the recursive algorithm in \cite{FLPR99}
we obtain a BP computation with $f(r)=O(1)$ and $L(r)=O(1)$.

{\bf Strassen.}\footnote{We correct
a typo in the cache bound for Strassen found in many papers, starting
with \cite{FLPR99}.}
We expose the parallelism in Strassen's matrix multiplication algorithm that
multiplies two $n\times n$ matrices by recursively multiplying seven
$n/2 \times n/2$ matrices, and
performs the matrix additions for the divide and combine steps
using MA.
This results in an HBP computation
that is of type 2, with $c=1$ collection of
$v=7$ subproblems of size $s(m) = m/4$,
where $m=n^2$ is the size of the matrix.
This algorithm computes the 7
recursive submatrices in
new subarrays.
These matrices are then combined with matrix additions and subtractions
(performed using MA) according to Strassen's algorithm, and the final
four submatrices are written back to the four quadrants in the parent matrix.
Thus each variable in this algorithm
is written only a constant number of times, and the algorithm is
inherently limited access.
When the matrices are in the BI layout, this computation has
$f(r)=O(1)$ and $L(r)=O(1)$.
The sequential cache complexity is $\Theta(\frac{n^{\lambda}}{B M^{\gamma}})$,
where $\lambda = \log_2 7$ and $\gamma = (\lambda/2)-1$.

\vhalf
Since we have assumed in the above algorithms that matrices are in the
BI layout, we need methods to convert between the traditional RM (row major)
layout and the BI layout. It turns out that
RM to BI is easy to execute
with $O(1)$ block-sharing, while BI to RM requires more effort.

\vhalf
{\bf RM to BI.}
We use a simple BP computation that recursively
converts each quadrant in parallel, with all writes in BI order.
The writes are thereby arranged so that tasks share $L(r)=O(1)$ blocks
for writing.
Reading, however, is only $f(r)= \sqrt{r}$-friendly. This is a BP
computation, so it is a type 1 HBP.

\vhalf
By employing
RM to BI initially
and suitable versions of BI to RM conversion
at the end (described below),
we obtain algorithms RM-MT
(use BI-RM (gap RM)),
and RM-Strassen (use BI-RM for FFT).
We now describe
several different methods for converting from BI to RM.

\vhalf
{\bf Direct BI to RM.}
This simple method uses the
same recursion as the direct RM to BI method mentioned above.
However, since the writes are to an output matrix in RM,
both $L(r)$ and $f(r)$ are $\sqrt r$.

\vhalf
We now present two improved algorithms for this (with respect to block misses),
of which only the first method performs $O(n^2)$ work.

\noindent
1.
{\bf BI-RM (gap RM).}
This is an $O(\log n )$ parallel running time, $O(n^2)$ operation
algorithm.

This is the same as Direct BI to RM, but to mitigate the block miss
cost, we use a \emph{gapping} technique.
The destination array representing the RM matrix
will be given gaps as follows:
between $r\times r$ subarrays (for values of
$r$  corresponding to  recursive subproblems) the rows will be given a length
$r/\log^2 r$ gap.
Now, tasks of size $r^2$ for $r=\Omega(B\log^2 B)$ share zero blocks for their writing.
This gives a cost of
$O(Br)$
 for the block misses for a size $r^2$ task, for
$r=O(B\log^2 B)$.
So $L(r^2)=O(r)$, but only for $r\le B\log^2 B$.

The justification for this choice of size is that
it only increases the size of the array by a constant multiplicative factor
  (for $\sum_{r=2^i}\frac{1}{ \log^2 r} =O(1)$).

Indeed a gap of $r/[\log r(\log\log r)^2 ] $,
or any analogous sequence of iterates, also works,
reducing the block miss cost correspondingly.

Having written to an array with gaps
one needs to compress the array using a standard scan.
This is a BP computation which
has $f(r)=O(1)$ and $L(r)=O(1)$.

\vhalf
\noindent
2.
{\bf BI-RM for FFT}.
This is an $O(\log n)$ parallel running time,
$O(n^2\log\log n)$ operation algorithm.
The algorithm divides the input
BI array of length $n^2$ into $n$ subproblems,
each of which it recursively converts to the RM order.
Then, using a BP computation, it copies the $n$ subarrays into one
subarray, accessing data according to the RM order in the target output.
This is a type 2 HBP computation that calls $c=1$ collection of
$v(n^2)=n$ subproblems of size $s(n^2) = n$.
The BP computation for the copying
 is organized so that the writes are in RM order, and hence
$L(r)=O(1)$.

We now show that $f(r)=O(\sqrt{r})$, assuming $M \geq B^2$.
Consider a size $r$ task $\tau$ performing a portion of the computation on a
$k \times k$ subproblem. The input to this $k \times k$ subproblem
consists of $\sqrt k$ rows of $\sqrt k$ submatrices that
are $\sqrt k \times \sqrt k$. Each of these $\sqrt k \times \sqrt k$
matrices has
already been converted to RM by recursive calls.

Let $r=s \cdot k + s' \cdot \sqrt k + s''$,
where $0 \le s',s'' < \sqrt{k}$.
We consider here the case
when $s \geq 1$;
the case when $s=0$ is handled similarly.

The task $\tau$ reads in
$s$ full rows of the output $k \times k$ matrix, plus part of the
$(s+1)$st row.

If $B > \sqrt k$ then by the tall cache assumption, $M \geq B^2 > k$. Hence,
when the data is read row by row according to the output $k\times k$
matrix, one block will be read from each of the $\sqrt k\times \sqrt k$
matrices in a given row. Further, under LRU, the $\sqrt k$
blocks will be all in
cache when the data for the next row of the output matrix is read, and
hence, within each $\sqrt k \times \sqrt k$ matrix, the number of blocks
read is the scan bound, leading to a bound of
$\leq \frac{(s+1)\sqrt k}{B} + \sqrt k$ cache misses for this computation.
Hence the number of
cache misses is $O(r/B + \sqrt k)$.

If $B <\sqrt k$, then reading a single row in one of the
$\sqrt k \times \sqrt k$
matrices
will incur $O(\sqrt k/B)$ cache misses,
and hence the total number of cache misses is $O(r/B)$.

\vone
{\bf FFT.}
We expose the parallelism in the
six-step variant of the FFT algorithm
 \cite{Ba90,VS94}
which is shown to have optimal $Q(n,M,B)$
in \cite{FLPR99}. As noted in \cite{CSBR09}, this
is also a low-depth multicore algorithm. The algorithm views
the input as a square matrix, which it transposes, then performs
a sequence of two recursive FFT computations on independent parallel
subproblems of size $\Theta ({\sqrt n})$, and finally performs MT
on the result.
The sequential time is $O(n \log n)$ and the
sequential cache complexity
is $O(\frac{n}{B} \cdot \log_M n)$ \cite{FLPR99}, and the parallel
depth is readily
seen to be $O(\log n \cdot \log\log n)$.

We keep the matrices in the
BI representation.
Thus, the HBP algorithm FFT, when called on an input of length $n$,
makes a sequence of
$c=2$ calls to FFT on $v(n)=\sqrt n$ subproblems of size $s(n) = \sqrt n$
with a constant number of BP computations (mainly MT) performed
before and after each recursive call.
We have
$f(r)=O(1)$ and $L(r)=O(1)$,
outside of the cost to convert between BI and RM formats.
At the end, to convert to the RM format
we use BI-RM for FFT
Thus $f(r) = \sqrt r$ for the overall
FFT algorithm due to the need to use RM to BI and
BI-RM for FFT.

\vone
{\bf List Ranking (LR).}
The list ranking problem is known to require
$\min \{ n , \frac{n}{B} \log_M n\}$
cache misses even in a sequential computation.
We match the second term, since it
determines the bound for the normal range of parameter values.
Our algorithm for LR
uses the resource oblivious
sorting algorithm SPMS in our 
recent
paper \cite{CR10b},
whose bounds are the same as those for FFT.
The Euler tour and tree computation algorithms have
the same complexity as LR.

Efficient multicore algorithms for LR based on eliminating large independent
sets are given in \cite{AGS09,CSBR09,BGS09}.
As in \cite{BGS09} we adapt the PRAM algorithm
that performs $O(\log\log n)$ stages
of eliminating a constant fraction of the elements in the linked
list, and then switches to the basic pointer jumping algorithm
when the size of the linked list falls below $n/\log n$.
To find
a large independent set, we use the method MO-IS in \cite{CSBR09}
that constructs an $O(\log^{(k)} r)$-size coloring
of the linked list, and then extracts an independent set of size
at least $r/3$ (where $r$ is the current length of the linked
list) by examining elements of
each color class in turn.
A phase on a list of length $r$ performs $O(\log^{(k)} r)$ calls to
SPMS on inputs whose combined length is $r$.
Thus it can be shown
\cite{CSBR09} that
a phase on a list of length $r$ completes with
$O(\frac{r}{B} \log_M r)$
cache misses sequentially, and in
$O(\log r \cdot \log\log r \cdot \log^{(k)} r)$ parallel
time when using SPMS for sorting, since SPMS has parallel time complexity
$O(\log r \cdot \log\log r)$.
Once the algorithm switches to pointer jumping, each pointer jumping
stage can be performed with
$O(1)$ calls to SPMS and scans on
a list of length $O(n/\log n)$, hence
overall this incurs $O(n\log n)$ work and has
$O((n/B) \log_Mn)$ cache misses and
$O(\log^2 n \log\log n)$ parallel time. This
pointer jumping  phase of the algorithm gives rise to the dominant
cost for the basic parameters of LR listed in Table 1.

Since LR makes calls to the type 2 HBP sorting algorithm SPMS before calling
itself recursively on $c=1$ sequence of $v(n)=1$ subproblem of size
$s(n) \leq 2r/3$, this is a type 3 HBP algorithm.

To reduce the number
of block misses in the recursive calls, we introduce gaps between the
elements of the contracted linked list as follows: When the list has size
$n/x^2$, it is written in space $n/x$, using every $x$th location only.
Thus, when the list has size $n/B^2$ or less, no more block misses occur.
Note that this modification of the list ranking algorithm does not affect
the cache miss cost beyond a constant factor.
This is because each of the BP computations from which the sorting algorithm
SPMS is built has a cache miss cost no larger than the cost on an array
of equal length with all entries occupied. Hence this holds for SPMS as
well, and hence for each recursive call in the list ranking algorithm.
As the space used is still shrinking by a factor of at least
2 every 4 iterations, the same
geometrically shrinking costs occur, leading to the same asymptotic
bounds.

\vhalf
{\bf CC.}
We use the connected components algorithm in \cite{CSBR09}.
The dominant cost in this algorithm is
$\log n$ stages of list ranking, so our resource-oblivious
implementation increases
each of the work, parallel time,
cache complexity, and block miss cost
by a factor of $\log n$.

\subsection{Data Layout and Block Wait Costs}\label{sec:layout}

By definition, an HBP algorithm is a limited-access computation. Additionally,
all of the HBP algorithms we have considered are
are either $O(1)$-block sharing, or incorporate gapping.
These features are useful in reducing block wait costs. However,
space gets reused on the execution stacks of the cores, and this may
cause block misses beyond those that can be inferred by analyzing accesses
to variables in the algorithm.

\vhalf
\noindent
{\bf Execution Stacks.}
Each core $C$, when it
starts executing
a task $\tau$, will create an execution stack
$S_{\tau}$ to keep track of the procedure calls and variables
in the work it performs on $\tau$.
The variables on $S_{\tau}$ may be accessed by stolen subtasks also.
As $S_{\tau}$ grows and shrinks it may use and then stop using a block $\beta$
repeatedly. Thus, it could be the case that
a computation with limited-access and $O(1)$-block sharing still incurs
a large block wait cost due to accesses to the execution stacks.

Lower and upper bounds on the space used for the variables declared
in an HBP procedure affect the block wait
as shown in our companion paper \cite{CR11}.
To obtain our bounds, we need that the space used by the local variables of a Type 2 HBP
procedure dominate the logarithmic space used by its BP subtasks.
For simplicity in our presentation, we make this a requirement that linear space be used
by these local variables.
All of our algorithms satisfy this property.

Accordingly, we make the following definition.

\begin{definition}
\label{def:spc-bdds}
A Type 2 HBP algorithm is \emph{Exactly Linear Space Bounded} if the variables declared
by each Type 2 task $\tau$ use space $\Theta(|\tau|)$.
\end{definition}

In our companion paper \cite{CR11} we prove the following lemma.

\begin{lemma}\label{lem:HBP-lim-access}
(i)
Let $\cal A$ be a limited-access BP Algorithm and let
$\tau$ be either the original task in the computation of
$\cal A$ or a task which is stolen during the
execution of $\cal A$.
Let $\beta$
be a block used for
$\tau$'s execution stack $S_{\tau}$.
Then $\beta$ incurs a block delay of $O(\min\{B, \log(|\tau|)\})$
during  $\tau$'s execution.

(ii)
Let $\cal A$ be a limited-access, exactly linear space bounded Type 2 HBP algorithm,
for which each collection of
recursive calls
consists of subproblems of size at most
$s(n) \le (1 - \gamma)n$,
for some constant $0<\gamma<1$.
Let $s^{(i)} (n)$ be the function $s$ iterated $i$ times.
Let $\tau$ be either the original task in the computation of
$\cal A$ or a task which is stolen during the
execution of $\cal A$.
Let $\beta$
be a block used for
$\tau$'s execution stack $S_{\tau}$.
Then the number of transfers of block $\beta$
during the execution of $\tau$
is bounded by
\begin{eqnarray*}
Y(|\tau|, B) = \left\{
\begin{array}{ll}
O(cB) & \text{if}~s(|\tau|) \ge B \\
 O( \sum_{i \ge 0}  c^{i}\cdot s^{(i)}(|\tau|)) & \text{otherwise}
\end{array}
\right.
\end{eqnarray*}
If $s(n)\le (1-\gamma)n/c$ this is an $O(\min\{cB, |\tau|\})$ bound.
\end{lemma}

All of our HBP algorithms satisfy the requirements of the above lemma.

\vhalf
\noindent
{\bf Data Layout in a BP Computation.}
The computation at a node $u$
during the up-pass in a BP computation
involves updating $O(1)$ data on
its execution stack, spread across at most $c=O(1)$ blocks,
and possibly $O(1)$ updates to the output data.
In the case of output data we assume that the data for each
size $r$ BP computation
is in an array of size $O(r)$,
and is stored according to an in-order traversal of its
up-tree. So, for instance, in M-Sum, the output data at each node
in the up-tree is the value of the sum of the input values at the leaves
of the node's subtree, and these values are stored in the order of an
in-order traversal of the up-tree.
The advantage of this layout is that it will result in no block misses for
accessing output data at levels in the BP tree where the number of leaves
in each subtree is greater than $B$.

\section{Priority Work Stealing Scheduler (PWS)}
\label{sec:pws}

In this section we present {\it PWS}, a deterministic work stealing scheduler.
We show that the excess cache misses
over the sequential cache miss cost
in a BP computation
when scheduled under PWS is $O(p M/B)$
with a standard tall cache $M\geq B^2$ if $f(r) = O(\sqrt r)$.
For block misses, we bound the
block wait cost for a
size $n$ BP computation with $L(r) = O(1)$ by $O(Bp \log B)$.
We build on these results to obtain bounds for our HBP algorithms.
In our analysis, we also bound other costs, including {\it usurpation
costs} and {\it idle time}.
In Section \ref{sec:pws-impl}
we give a distributed implementation of PWS.

PWS assigns to each task a priority that decreases with increasing depth
in the computation dag, so there are up to
$T_{\infty}$ different priorities.
Steals in PWS proceed in {\it rounds}, one for each priority, and
are performed in non-increasing priority order.

When a core $C$ executes a task $\tau$, the tasks it places on its task queue
will have lower priority (i.e., larger computation depth) than $\tau$, hence
it is not difficult to see that at most $p-1$ tasks of any given priority
are stolen under PWS.
Further, the priorities are assigned so that all tasks with a given
priority have the same size, to within a constant factor;
note that this is readily
accomplished for BP and balanced HBP computations.

In this section we bound the various costs incurred by
a balanced HBP computation
when scheduled under PWS. We establish that
the main costs arise from the cache and block
misses, which we bound in terms of their excess, which is the amount
by which these costs exceed the sequential cache complexity of the
computations.
The following two lemmas give the bounds we derive for
the cache and block miss excess for the
Type 2 HBP algorithms we consider.
We analyze LR (Type 3) and CC (Type 4) separately, building on these results.
These lemmas are established in Sections \ref{sec:pws-cache} and
\ref{sec:block-pws} respectively.

Given a function $s(n) < n$, recall that
$s^*(n)$ is the minimum $i$ such that $s^{(i)}(n) \leq c$, for a suitable
constant $c$; here, $s^{(0)}(n)=n$, and $s^{(i)}(n) = s(s^{(i-1)}(n))$,
if $i>0$.
We also define $s^*(n,M)$ as the minimum $i$ such that $s^{(i)}(n) \leq M$.

\begin{lemma}
\label{lem:c-miss-rec-perf}
Let $\Pi$ be a balanced Type 2 HBP computation of size $n \geq Mp$,
and let $c,~ s(n),$
and $f(r)$ be as defined earlier. Then, the cache miss excess for $\Pi$
when scheduled  under PWS has the following bounds with a tall cache
$M \geq B^2$.
\vhalf

(i)  If $c=1$, $f(r)=O(\sqrt{r})$: $O(p\frac{M}{B}s^*(n,M))$.

(ii) If $c=2$, $f(r)=O(\sqrt{r})$, and $s(n)=\sqrt{n}$:
$O(p\frac{M}{B} \frac{\log n}{\log M})$.

(iii) If $c=2$, $f(r)=O(\sqrt{r})$, and $s(n)=n/4$: $O(p[\frac{\sqrt{nM}}{B}+
    \frac{\sqrt{n}}{\sqrt{M}} \sum_{i\ge 0} 2^i f(M/4^i) ])$.

\end{lemma}

\begin{lemma}
\label{lem:block-miss-rec-perf}
Let $\Pi$ be a balanced Type 2 HBP computation of size $n \geq Mp$
with $\alpha = 1/2$ and $L(r) = O(1)$, which is exactly linear space bounded,
and let $c,~ s(n),$
and $L(r)$ be as defined earlier. Then, the block miss excess for $\Pi$
when scheduled  under PWS has the following bounds.

\vhalf
(i) $c=1$: a cost of
$O(pB\log B \cdot s^*(n))$ cache misses.

(ii) $c=2$ and $s(n) = \sqrt{n}$: a cost of
$O(pB\log n \log\log B)$
cache misses.

(iii) $c=2$ and $s(n) = n/4$: a cost of
$O(pB \sqrt{n})$
cache misses.
\end{lemma}

\subsection{PWS Scheduling}

As mentioned earlier, the
PWS scheduling requires tasks to have integer priorities with
the property that on any root to leaf path in the computation tree $T$
the priorities are strictly decreasing.
PWS proceeds in {\it rounds}, one for each integer priority.
The steals under PWS are performed in decreasing priority order, which,
loosely speaking, is also a size-based breadth-first search order
(since
the priorities will be a function of the sizes).

The first round starts when the core $C$ that began
computation $\Pi$ places its first task $\tau_1$ on its task queue.
Let this task have priority $d_1$.
The priority of round 1 is $d_1$, and during this first
round, any of the other cores can steal $\tau_1$. Round 1 concludes
when $\tau_1$ has been removed from the head of $C$'s task queue and
assigned to an idle core $C'$.

 In general, a round with priority $d$ concludes when no task queue
has a task of priority $d$ at its head
and
every non-idle core has generated a task on its task queue.
This starts the next round whose
priority $d'$ is the priority of
the highest-priority task at the head of a task queue.
During this round, tasks of priority
$d'$ at the heads of task queues are stolen by idle cores until no
task at the head of a task queue has priority $d'$.

\begin{observation}\label{ob1}
The priorities
of tasks in the task queue
of a core at any point in time are strictly decreasing from top
(i.e., the head) to bottom.
\end{observation}

\begin{observation}\label{ob2}
If a steal request by a core is unsuccessful in a round with
priority $d$,
then any remaining task has smaller priority.
\end{observation}

\begin{observation}\label{ob3}
For each task priority,
there are at most $p-1$ tasks of that priority that are stolen.
\end{observation}

\begin{corollary}\label{cor4}
Let the number of distinct task priorities be $D'$.
The total
number of steal attempts (including both
successful and unsuccessful steals) across all cores is at most
$2  \cdot p \cdot D'$.
\end{corollary}

\subsection{Cache Misses under PWS}\label{sec:pws-cache}

In a BP computation, the priority of a node at depth $i$ is $i$, and
priorities decrease with increasing depth.
By the definition of a BP computation, all tasks with a given
priority have the same size, to within a constant factor.

We start with the following useful lemma, which uses $c_1, ~c_2$ from
the definition of a BP computation.

\begin{lemma}\label{o4}
Consider the downpass of
an $f$-cache friendly
BP
computation $\Pi$ scheduled under PWS on $p$ cores,
each with a cache of size $M$.
Let $d$ be the
number of distinct priorities in
$\Pi$,
and
$Q$ its sequential cache complexity.

If $f(r)= O(r/B)$ for
$r \geq (2c_1/c_2) \cdot M$,
then the number of additional cache misses incurred by all
stolen tasks of size greater than $2M$
is $O(Q)$.
\end{lemma}
\begin{proof}
By Lemma \ref{lem:c-miss-basic-bdd} and since $c_1 \leq c_2$,
a stolen task of size greater than $2M$ has zero cache miss excess.
What remains to be argued is that
for each task $\tau$ of size greater than $2M$,
the additional cache miss cost of the
task kernel $p_{\tau}$
that remains after tasks
of size $2M$ or larger are stolen at a core
can also be bounded.
For this, consider a task $\tau$ (either the root task
or a stolen task) executing at a core $C$. If there are no steals then
the result holds vacuously.
Otherwise let $\tau_v$ be the last task
of size $2M$ or larger
stolen from $C$ while
executing $\tau$,
and let $v$ be
the node in the computation tree for
$\tau$ corresponding to $\tau_v$.
Then, by Observation~\ref{obs:stolen-tasks},
the task $\tau_w$ at the sibling of $v$ is not a stolen task and is
part of the task kernel $p_{\tau}$.
But $\tau_w$ has
size at least $(c_1/c_2) \cdot |\tau_v| \geq (2c_1/c_2) \cdot M$.
Further, $p_{\tau}$ consists of a sequence of maximal sub-tasks in $\tau$,
all of which have size at least
$2\frac{c_2}{c_1}M$
(since $\tau_w$ is the last and smallest of these tasks).
Hence since $f(r) = O(r/B)$ for
$r\geq (2c_1/c_2) \cdot M$,
the sum of the costs of $f(r_i)$,
where $r_i$ is the size of the $i$th maximal task in $p_{\tau}$,
is bounded by $p_{\tau}$'s
sequential cache complexity.

The $O(1)$ cache miss cost at each task-head at the parent
of a stolen task can be bounded as follows. Consider the tree formed by
the nodes for stolen tasks of size $2M$ or larger,
the ancestors of these nodes, and all siblings of
such nodes and their ancestors.
This tree is defined with respect to the tasks computed at all the cores.
This is a binary tree in which each non-leaf has two children, and each
leaf has size at least
$(2c_1/c_2) \cdot M$.
Thus each of these leaves incurs $\Omega(M/B)$ cache miss cost.
It follows that each leaf can safely be allocated a further $O(1)$ cache misses.
As there are more leaves than internal nodes, the $O(1)$ cache-miss cost
for each each internal node can be allocated to a distinct leaf;
this includes the nodes for each task-head at the parent of a stolen task.

Hence by
the observation that any sequential execution of
a task of size $r\ge 2M$ will incur at least
$r/B$ cache misses,
we can bound the cache
miss cost of executing the
task kernels by the
sequential cache complexity $O(Q)$.
\end{proof}

The above lemma bounds the additional cache misses incurred when a task
is stolen.
We note that there is another source for
cache misses incurred due to steals, and this occurs when control of a task
transfers to a stolen task after a join in the computation. This motivates
the following
definition of {\it usurpation}.

\begin{definition}\label{def:usurp}
Let $\tau$ be a task whose kernel is being executed by core $C$. Let
$C$ complete its execution of a subtask $\tau''$ in $\tau$'s kernel,
and suppose that the join is
performed after $\tau''$ by its sibling task $\tau'$, which is a stolen
subtask executed by another core $C'$. Then, $C'$ will take over the remainder
of the execution of $\tau$'s kernel. This change in the core executing
$\tau$'s kernel is called a \emph{usurpation} of $\tau$ by $C'$.
\end{definition}

A usurpation of $\tau$'s kernel by core $C'$ could result in additional
cache misses over the sequential computation. We will compute this cost
separately in Section \ref{sec:hbp-cache}. For the remainder of the current
section, we will bound the cache miss excess due to steals as reflected
in Lemma \ref{o4}.
We start by  bounding
the number of cache misses incurred by all stolen tasks in
a BP computation.
%

\begin{lemma}\label{lem:pws2}
 Let $\Pi$ be an
$f$-cache friendly BP computation of size $n$, and consider its
execution on $p$ cores, each with a cache of size $M$.
In addition,
let $Q$ be the number of cache misses in a sequential execution of
$\Pi$.
Finally, suppose that $f(r)= O(r/B)$ for
$r \geq (2c_1/c_2) \cdot M$.
Then, when executed using PWS,

\noindent
(i) The number of cache misses is bounded by

\vspace{-.1in}
\[
O(Q + p(\frac{M}{B} + \log B) + \sum_{j\ge 0} p \cdot
f(2M\alpha^{j}) ).
\]
$(ii)$
If $M \geq B \log B$ and $f(r) =O(1)$,  the number of cache misses
is bounded by
  $O( Q + p \cdot  M/B ~)$.

\noindent
 $(iii)$
If $M\ge B^2$ and $f(r) =O(\sqrt{r})$,  the number of cache misses
is bounded by
  $O( Q + p \cdot  M/B ~)$.
\end{lemma}
\begin{proof}
$(i)$
By Lemma \ref{o4},
it suffices to bound the cache misses
incurred by stolen tasks of size less than $2M$.
Let $d$ be the largest priority of any such task.
Then tasks of priority $d-j$ have size at most $2M\alpha^j$.
Each such task
incurs $O(\ceil{M\alpha^j/B} +
f(2M\alpha^j))$
cache misses. Summing
over all tasks yields $O(p(M/B + \log B) + p\sum_{j\ge 0}
f(2M\alpha^j))$
cache misses.

$(ii)$ The summation is $O(pM/B)$ in this case, and the bound follows.

$(iii)$
Since
$f(r) =O(\sqrt{r})$, for
$r \ge M$
we have
$f(r)=O(r/\sqrt{M})=O(r/B)$.
Thus part (i) applies.
It remains to observe that
$p\sum_{j\ge 0} f(2M\alpha^j)\le
O(p\sqrt{M})=O(p\cdot M/B)$,
which yields the claimed bound.
\end{proof}

Let $Q(\pi)$ be the sequential cache complexity of a computation
$\pi$, and let $Q_{PWS}(\pi)$ be the number of cache misses incurred by
$\pi$ when it is scheduled under PWS.
Then define the {\it PWS cache miss excess},
$Q_C (\pi)$, as follows:
$Q_C (\pi) = Q_{PWS}(\pi)$ if $Q_{PWS}(\pi) = \omega (Q(\pi))$, and
$Q_C (\pi) = 0$
if $Q_{PWS}(\pi) = O(Q(\pi))$.
Thus for instance, in
parts $(ii)$ and $(iii)$
of the above lemma,
the PWS cache miss excess for any BP computation with
$f(r)= O(\sqrt r)$ and $M \geq B^2$ is $Q_C (\pi) = O(Mp/B)$.
In particular, when $n \geq Mp$, i.e., when the input does not fit into
the caches, the PWS cache miss excess is zero.

The following corollary addresses smaller input sizes ($n < Mp$)
for a case that arises in our list ranking and
graph algorithms.

\begin{corollary}\label{cor-small}
Let $\pi$ be an $f$-friendly BP computation
of size $n<Mp$ and suppose that $\alpha=1/2$ and $f(r)=O(1)$. Then, the
PWS cache miss excess
$Q_C (\pi)$
is:

(i) For $Bp \leq n < Mp$,
$~~Q_C (\pi) =  O\left(p \log B + \frac{n}{B} \cdot \log \frac{4pM}{n}\right)$.

(ii) For $p \leq n < Bp$,
$~~Q_C (\pi) = O(\frac{n}{B} \log 2\ceil{\frac{ \min\{n,M\} }{B}}   + p\log\frac{2n}{p}~)$.

(iii) For $n<p$,
$~~Q'' = O(\frac{n}{B} \log 2\ceil{\frac{ \min\{n,M\} }{B}}   + n)$.
\end{corollary}
\begin{proof}
We proceed as in the proof of Lemma
\ref{lem:pws2}.
As before, we need to sum the cache misses due to stolen
tasks of size smaller than $2M$.
There are $\min\{p,\frac{n}{2M\alpha^j}\}$ tasks of size
$2M\alpha^j$,
each generating
$O(\ceil{ \frac{2M\alpha^j}{B} })$
cache misses.
Summed over all $j$,  in Case (i) this gives
$O(p \log B + \frac{n}{B} \cdot \log \frac{4pM}{n})$
cache misses, in Case (ii),
$O(\frac{n}{B} \log \frac{2 \min\{n,M\} }{B}  + p\log\frac{2n}{p}~)$
cache misses,
and in Case (iii),
$O(\frac{n}{B} \log \frac{2 \min\{n,M\} }{B}  + n)$
cache misses.
We argue only Case $(i)$.
Summing over the increasing values of $j$, starting at $j=0$,
there are terms $O(n/B)$ for $\ceil{ (\log p)/(n/2M)}$ values of $j$, followed
by terms $p\ceil{\frac{2M\alpha^j}{B}}$ up to the first value of $j$
for which $M\alpha^j \le B$, which total $O(n/B)$,
followed by terms $p\cdot O(1)$ for $\log B$ values of $j$.
The bound in (i) now follows readily.
\end{proof}

\vone

A {\it BP collection} is a collection of parallel independent BP
computations; similarly, an
{\it HBP collection} is a collection of parallel independent HBP
computations. Such collections are generated when parallel recursive
calls are made in an HBP computation.
 The {\it size} of a BP or HBP collection is the maximum
size among the independent computations in the collection
and its \emph{total size} is the sum of the sizes of the
independent computations in the collection.
It follows from the definition of a balanced HBP computation that
in any BP or HBP collection that occurs in it,
every independent computation in the collection will have the same
size, to within a constant factor (or $c_2/c_1$).

As our cache miss bounds for a single BP computation
depend only on the binary forking
and priorities based on size, the previous bounds apply unchanged to a
BP collection, yielding the following corollary.

\begin{corollary}
\label{cor:indptBP}
Let $\Pi'$ be an $f$-friendly BP collection of size $r$,
whose total size is $n \geq Mp$.
Then the
PWS cache miss excess for $\Pi'$ is
bounded by
\[
O\left(p \cdot (~ \ceil{\frac{\min\{M,r\}}{B}} + \log \min\{r,B\} + f(r)~)\right).
\]
\end{corollary}
\begin{proof}
The argument is similar to that of Lemma \ref{lem:pws2}.
The one change is that $f(r)$ cannot be bounded by $r/B$.
\end{proof}

\vone
\noindent
{\bf Cache Misses in the Up-pass.}
In the {\it up-pass} that follows a down-pass,
the computation involving the activation of a suspended
task $\tau$ follows the mirror image of the initial pass that forked
the children of that suspended task.
Since we have assumed
that there is only $O(1)$ computation at each suspended task-head,
the PWS cache miss
cost of the up-pass is readily bounded.
The main additional cost
arises because a stolen task may assume
the remaining work of its parent
due to usurpation,
in which case it needs to read data
from its parent's local stack.
The cost is bounded by the total size of these stacks divided by $B$, which is
$O(\mbox{number of nodes in the BP tree}/B)$, plus an additional
$O(1)$ cache misses per
steal. This is subsumed by the cost of the downpass.
Thus there is zero contribution from the up-pass of a BP
computation $\pi$ to the PWS cache miss excess for $\pi$.
Note, however, that these reads may interleave with writes to these
same execution stacks by other cores,
and we will analyze their costs as part of the
block miss analysis.

\subsubsection{HBP computations}
\label{sec:hbp-cache}

We extend priorities to a balanced HBP computation $C$ in a natural way.
Priorities decrease with depth, and in each BP collection in $C$,
all nodes at depth $i$ in the collection have the same priority, $i \geq 0$.
Thus, all nodes with the same priority will have the same size, to within
a factor of $c_2^2/c_1^2$.

We now bound the cache miss excess in the HBP computations we consider.
First,
by the same argument as for Lemma \ref{o4} we obtain:

\begin{lemma}\label{lem:HBP-cache-large}
Consider
an $f$-cache friendly balanced HBP
computation $\Pi$ scheduled under PWS on $p$ cores,
each with a cache of size $M$.
If $f(r)= O(r/B)$ for
$r \geq (2c_1/c_2) \cdot M$,
there is zero PWS cache miss excess due to
stolen tasks of size $2M$ or more.
\end{lemma}

\vhalf
\noindent
{\bf Usurpations.}
Consider two
successive HBP collections $H_1 =\{h_{11}, h_{12}, \cdots, h_{1k} \}$ and
$H_2= \{h_{21}, h_{22}, \cdots , h_{2k}\}$, where, for each $i$,
 $h_{2i}$ follows $h_{1i}$ in the sequential computation.
Let $h_{1i}$, for some $i$, start its computation on
some core $C$.
Because of usurpation it may be that a core $C'$, that
stole a subtask from $C$ within $h_{1i}$, will be the one starting the
computation $h_{2i}$; we call $C'$ a \emph{usurper}. Further, even if
$C$ starts the computation of $h_{2i}$, it may be that, due to steals,
$C$ has not read
in some of the data for $h_{1i}$ that potentially gets reused in $h_{2i}$.
In this case we say that $C$ is \emph{semi-usurped}.
In both cases (usurped and semi-usurped) we need to bound the cost of
the additional reads needed to execute $h_{2i}$ since our analysis in
Lemma \ref{lem:pws2} assumed that the core starting the BP computation had
the same state as
in the sequential computation.
If $k$ is large,
then conceivably there could be a large number of usurpers at $H_2$ resulting
in a large cache miss cost. We
now argue that this is not the case for balanced HBP collections.

\begin{lemma}\label{lem:usurpers}
Let
$H_1 =\{h_{11}, h_{12}, \cdots, h_{1k} \}$ and
$H_2= \{h_{21}, h_{22}, \cdots h_{2k}\}$
be successive balanced HBP computations.
Then, there are at most $p-1$ usurpers and
semi-usurpers.
\end{lemma}
\begin{proof}
Let the first sub-task stolen in $H_1$ be $h'$, and let it be a sub-task
of $h_{1j}$, for some $j$. At the time $h'$ is stolen, all of the
$h_{1i}$ have started their execution, since otherwise there would
be a task of higher priority than $h'$ available for stealing.
But then, at most $p-1$ of the $h_{1i}$ can still be in the process
of being executed. These are the only tasks that can be usurped or
semi-usurped.
\end{proof}

 As noted earlier, all of our HBP algorithms are balanced, hence the
above lemma applies to them. The sorting algorithm SPMS in \cite{CR10b}
is not a balanced HBP computation, however a different argument bounds
the cost of usurpations in that algorithm.

\vone
\noindent
{\bf Cache Misses in Balanced HBP Computations.}
By Lemma \ref{lem:HBP-cache-large}, stolen tasks of size $2M$ or larger
in an HBP computation incur no cache miss excess under PWS.
In the following lemma, we
bound the excess due to steals of smaller tasks in an HBP
computation using Lemma \ref{lem:pws2} and Corollary \ref{cor:indptBP}.
and we bound the cost of usurpations using Lemma \ref{lem:usurpers}.

\begin{lemma}\label{lem:HBP-c-miss-total-cost}
Let $\pi$ be an $f$-friendly balanced HBP computation with $f(r) =O(\sqrt{r})$.
If $M\ge B^2$, the
PWS cache miss excess for $\pi$ is
\vspace{-.1in}
\[
O(p \cdot v' (M/B +\log B)
+p \cdot \sum_i (n_i/B +\log B + f(n_i)) +p\cdot \sum_j (\log n_j + f(n_j))~) ,
\]
where
$v'$ is the number of BP collections of size at least $2M$,
the first sum is over BP collections whose
size is $n_i$, $B<n_i < 2M$,
and the second sum is over BP collections whose size is $n_j < B$.
\end{lemma}

\begin{proof}
The cost of stealing small tasks is bounded by applying
Lemma \ref{lem:pws2} to the initial computation, and then applying
Corollary \ref{cor:indptBP} to the BP collections in the computation.

To bound the cost of usurpers we apply Lemma \ref{lem:usurpers}.
Since there are at most $p-1$ usurpers and semi-usurpers for each
BP collection, and we have already charged for $p-1$ steals at the
start level for
this collection, the usurpations increase this cost by at most a
factor of 2 for each collection.
\end{proof}

Using Lemma \ref{lem:HBP-c-miss-total-cost}, we obtain the following
Lemma \ref{lem:c-miss-rec-perf},
which was stated at the beginning of Section \ref{sec:pws}.

\noindent
{\bf Lemma \ref{lem:c-miss-rec-perf}.}
{\it
Let $\Pi$ be a balanced Type 2 HBP computation of size $n \geq Mp$,
and let $c,~ s(n),$
and $f(r)$ be as defined earlier. Then, the cache miss excess for $\Pi$
when scheduled  under PWS has the following bounds with a tall cache
$M \geq B^2$.

(i)  If $c=1$, $f(r)=O(\sqrt{r})$: $O(p\frac{M}{B}s^*(n,M))$.

(ii) If $c=2$, $f(r)=O(\sqrt{r})$, and $s(n)=\sqrt{n}$:
$O(p\frac{M}{B} \frac{\log n}{\log M})$.

(iii) If $c=2$, $f(r)=O(\sqrt{r})$, and $s(n)=n/4$: $O(p[\frac{\sqrt{nM}}{B}+
    \frac{\sqrt{n}}{\sqrt{M}} \sum_{i\ge 0} 2^i f(M/4^i) ])$.

}

\subsection{Block Misses Under PWS}
 \label{sec:block-pws}

Let $\pi$ be a
BP collection of size $r$ and total size $n$
in which each task $\tau$
shares at most $L(|\tau|)$ blocks with other tasks.
There are at most $\frac{c_2}{c_1}n/r$
BP trees of size $r$ in the collection, hence
each level below level $\log (pr/n)$ will have at least $p$ tasks.
Under PWS,
there are at most $p$ steals at any level,
hence the total number of shared blocks across all cores
due to steals under PWS in this computation is
$X(r)$, where
\begin{eqnarray}\label{eq:X}
X(r)\leq\sum_{\log (n/r) \le i \leq \log \min\{p,r^{1/\log 1/\alpha}\}} 2^i \cdot L(r\alpha^i) ~+~ p \sum_{\log p< i \le \log r^{1/\log 1/\alpha}} L(r\alpha^i)
\end{eqnarray}
Hence, using Lemma~\ref{lem:HBP-lim-access}, we obtain that the sum of the
block waits across all cores during this computation
is bounded by
$Z(r) = Y(B) \cdot X(r)$, for $r\ge B$, and by
$Z(r) = Y(r) \cdot X(r)$, for $r < B$.

For a computation $\pi$ scheduled under PWS, let $Q_{PWS,B}(\pi)$ be
the block wait cost of $\pi$, and as before, let $Q(\pi)$ be the sequential
cache complexity of $\pi$.
As with
the PWS cache miss excess, we
define the
{\it PWS block miss excess}, $Q_B(\pi)$,
as $\pi$'s block wait cost $Q_{PWS,B}(\pi)$
if $Q_{PWS,B}(\pi) = \omega (Q(\pi))$, and $Q_B(\pi)=0$ otherwise.

\begin{lemma}\label{lem:block}
Let $\pi$ be
the downpass of a BP collection of size $r$.
Suppose that $Y(|\tau|)= O(\min\{B, |\tau|\})$.

(i) If the block-sharing function $L(|\tau|)= O(1)$,
then,
for $r\ge B$,
the PWS block miss
excess for $\pi$ is $O(p\cdot B \log B)$;
for $r < B$, the block miss excess is $O(p\cdot r \log r)$.

(ii) If the block-sharing function $L(|\tau|) = \sqrt{ |\tau|}$,
and
$\alpha =\frac12$,
then,
for $r\ge B$,
the PWS
block miss excess for $\pi$ is $O(B \cdot \sqrt{pr})$;
for $r<B$, it is $O(r \cdot \sqrt{pr})$.
\end{lemma}
\begin{proof}
$(i)$
For $r \ge B^2$,
we first consider the stolen tasks of size $B^2$ or larger.
Similar to the proof of Lemma \ref{o4}, we can see that the
kernel of task $\tau$
(either a stolen subtask or an original task
in this BP collection)
after all its stolen subtasks of size $B^2$ or larger are removed
still has size $\Omega(B^2)$.
Consequently, its execution (which may include further steals of smaller
subtasks) will incur a cache miss cost of $\Omega(B)$.
As the block wait cost for a stolen task is $O(B)$,
we can conclude that the block
miss excess
for all original and stolen tasks of
size $B^2$ or larger
is zero.

If $r \geq B^2$, then the
remaining cost
is bounded by
\begin{eqnarray}
\label{eq:X-modified}
  B \cdot p\sum_{i>\log_{1/\alpha} r/B^2} L(r\alpha^i) = O(Bp\log B).
\end{eqnarray}
For $r < B^2$, the cost is bounded by
$O(\min\{r,B\} \cdot p \sum_{i\ge 0} L(r\alpha^i))
=O(p \min\{r,B\} \log r)$.

\vhalf
$(ii)$ This follows directly from Equation \ref{eq:X}.
\end{proof}

\vone
\noindent
{\bf Block Misses in the Up-pass.} Recall that there are no steals during the up-pass.
Instead, as mentioned in Section \ref{sec:comp-model}, at any non-leaf
node $u$ in a BP tree,
of the tasks at its two children, the later finishing one
will continue the up-pass from $u$ toward the root of its BP tree.
The following lemma assumes the data layout given in Section \ref{sec:layout}.

\begin{lemma}
\label{lem:up-block}
Let $\rho$ be the up-pass of a
BP collection of size $r$ and total size $n$, scheduled under PWS. Then,
for $r^{1/\log 1/\alpha} \ge p$,
the block miss excess for $\rho$ is
$O(\min\{r,B\}p (\log_{1/\alpha} r - \log (\max\{2,pr/n\})))$;
for $r^{1/\log 1/\alpha} < p$,
it is $O(\min\{r,B\}p)$.
For $\alpha = 1/2$,
and
$r \ge B$,
the block miss excess is $O(Bp \log B)$,
if $n\ge pM$ and $M \ge B^2$;
for
$\alpha=1/2$ and
$r < B$ it is $O(rp \log r)$.
\end{lemma}

\begin{proof}
We first note that
the computation at each node in the up-pass
is of constant size, hence by the limited-access property there is an
$O(\min\{r,B\})$ block wait cost at each node regardless of the nature of the
block-sharing function $L(|\tau|)$.

Recall that the two children of an internal node in
a BP tree have size
between $c_1 \alpha \cdot r$ and $c_2 \alpha \cdot r$, for some
$\alpha$ with $1/2 \leq \alpha <1$
and constants $0<c_1 \le 1 \le c_2$.
Hence the height of the tree is
$\log_{1/\alpha} r + O(1) = \log_2 r^y + O(1) $, where
$y = 1/\log_2 (1/\alpha)$.
Then, the number of nodes in the top level is at least $n/r$
and this number increases to $p$ by level
$l = \log q$, where $q = \max\{2, pr/n\}$.
(We assume that $M \ge c_2/c_1$ so that leaves are not encountered before
level $q$.)

We first consider the total block wait cost at the top
$l$ levels of the BP tree collection,
There are at most
$2p-1$ nodes in this portion of the
BP tree collection,
and the tasks at the two children of a node $u$ are the only ones
that access $u$ for their computation. By the limited access
property, each of these $4p-1$ tasks
incurs a block wait cost no more than
$O(\min\{r,B\})$.
Hence the contribution
of the top $\log p$ levels of the up-pass to the total block wait cost
is $O(\min\{r,B\}p)$.
For $q>r^y$,
this cost reduces to
$O(\min\{r,B\}r^y)$ over all the
$\log_{1/\alpha} r + O(1)$
levels of the up-pass.

Only in the case that
$q \leq r^y$
are there further levels to consider.
We now consider these bottom
$\log (r^y/q) + O(1)$
levels.
Under PWS,
there are at most
$p \log (r^y/q)$
stolen tasks in these levels.
Further, consider a block
$\beta$
on an execution stack
that is accessed by $x$ stolen tasks.
Each of these stolen tasks could incur a block wait cost of up to
$x$
due to its accesses to $\beta$.
But only one of these tasks would continue
computing further up its computation tree.
This is because the block $\beta$
contains the execution stack data for
a sequence of ancestors in the BP tree, and only the task
that completes the computation at the highest ancestor
will continue
the computation up the tree.
Hence, if there are $s$ different blocks accessed on the execution
stacks across all tasks during
this computation, and if $r_i$, $1\leq i \leq s$, is the number of
tasks accessing the $i$th block, then
$ 1 + \sum_{i=1}^s (r_i -1)$ is bounded by
$c_2/c_1$ times the total number of steals
in the
$\log (r^y/q)$ levels,
which is at most
$c p \log (r^y/q)$.
Since
$s=O(p \log (r^y/q))$ it follows that $\sum_{i=1}^s r_i = O(p \log (r^y/q))$,
hence by the limited access property,
the total block wait cost incurred in
these $\log (r^y/q)$ levels is $O(Bp\log (r^y/q))$.
The block wait cost for accessing other data items also
satisfies this bound since each task will access $O(1)$ data with
$O(\min\{r,B\})$
block wait cost,
which is accounted for in the above bound. Since $y = 1/\log_2 (1/\alpha)$,
this establishes that the block miss excess is
$O(\min\{r,B\}p (\log_{1/\alpha} r - \log (\max\{2,pr/n\})))$.

\vhalf
For $r \ge B$, when $\alpha = 1/2$,
the block miss excess is
$O(Bp \log n/p)$.

We now show that
$Bp\log (n/p)= O(n/B + Bp \log B)$
if
$n\ge pB^2$
(which holds if $n\ge pM$ and $M \ge B^2$).

\begin{eqnarray*}
Bp \log (n/p) &\leq&  Bp( \log (n/(pB^2)) + \log B^2 )\\
           &\leq& \frac{n/B}{n/(pB^2)} \cdot \log (n/(pB^2))  + Bp \log B^2
	   ~~\leq~~ n/B + 2B p \log B
\end{eqnarray*}

Hence the block miss excess for the up-pass is $O(Bp \log B)$ in this case.

For $r<B$, the claimed bound is immediate.
\end{proof}

\vone
We can now prove Lemma \ref{lem:block-miss-rec-perf} by applying
Lemmas \ref{lem:block} and \ref{lem:up-block}.

\vhalf
\noindent
{\bf Lemma \ref{lem:block-miss-rec-perf}.}
{\it
Let $\Pi$ be a balanced Type 2 HBP computation of size $n \geq Mp$
with $\alpha = 1/2$, which is exactly linear space bounded,
and let $c,~ s(n),$
and $L(r)$ be as defined earlier. Then, the block miss excess for $\Pi$
when scheduled  under PWS has the following bounds if $L(r) = O(1)$.

\vone
\noindent
(i) $c=1$: a cost of
$O(pB\log B \cdot s^*(n))$ cache misses.

(ii) $c=2$ and $s(n) = \sqrt{n}$: a cost of
$O(pB\log n \log\log B)$
cache misses.

(iii) $c=2$ and $s(n) = n/4$: a cost of
$O(pB \sqrt{n})$
cache misses.
}

\begin{proof}
The bounds are obtained by summing over the cost of the successive
collections of BP computations.

(i) follows from
Lemmas \ref{lem:block} and \ref{lem:up-block}
as there are
$s^*(n)$
successive
collections of BP computations
each collection costing
$O(Bp\log B)$ cache misses.

(ii) The cost of the BP computations is bounded by\\
$O(Bp\sum_{i=0}^{\log\log n-\log\log B} 2^i \log B$
 $ +p\sum_{j\ge 1} B
2^{j+\log\log n-\log\log B} \log B^{1/2^j})$
$=O(Bp\log n \log\log B)$.

(iii) The cost of the BP computations is bounded by\\
$O(Bp\sum_{i=0}^{\frac12 \log(n/B)} 2^i \log B$
  $+ \sum_{j\ge 1}
  B p\cdot 2^{j + \frac12 \log(n/B)} \log (B/2^{2j}))$
$= O(p \cdot \sqrt{nB} \log B + pB \sqrt n) =O(pB \sqrt{n})$.
\end{proof}

\subsection{Idle Time and Scheduling Costs}\label{sec:idle}

We now bound the total work
during a computation $\pi$
that could be attributed to idle
cores (i.e., when a core is neither computing, nor
waiting on a cache miss,
block miss,
or steal initiated by that core).

Idle time can occur when some cores are
executing the up-pass of a BP computation.
Consider the up-pass $\rho$ of a BP computation of size $n$.
We bounded the
block miss excess in Lemma \ref{lem:up-block}.
Here we consider the idle time incurred by such an up-pass. For this,
we need to
account for the time that some of the cores may need to wait in order for
the up-pass to complete so that new tasks can be generated on the task
queues and
become available for stealing.
This wait time starts at the time
$T$ when the last task in the downpass starts its computation, and it ends when
the computation completes at the root of the up-pass tree.

\begin{lemma}
\label{lem:up-idle}
Let $\rho$ be the up-pass of a limited-access BP
collection of total
size $n$.
The total idle time incurred during the up-pass is bounded by
$O(bp \cdot (\log n + B\log B))$,
where $b$ is an upper bound on the delay due to a cache miss.
\end{lemma}
\begin{proof}
Let $t$ be an upper bound on the
computation time at a leaf in the downpass of the BP computation, followed
by the computation from the leaf for
$\log (c_2B /c_1)$ levels up along
the path to the root of the up-tree.
We note
that $t = O(b B\log B)$ by the limited access property, and the fact that
base case tasks and up-pass nodes
perform $O(1)$ computation. Also, once the up-tree
computation is above the lowest
$l=\log (c_2 B /c_1)$
levels of the up-tree, a write
to each output data has
zero block wait cost,
since by
the in-order organization of the output data, the writes at any two
nodes are separated by at least a distance of $B$ in the output.
Thus, at levels above $l$
in the up-tree we only need to consider
block sharing on
the execution stacks of the tasks.

Let the number of blocks accessed at each node in the up-tree be
$c=O(1)$.
Let the last task in the downpass start its computation
at time $T$. We establish the following assertion by induction on
the height of a node $u$ in the BP tree.

\vhalf
\noindent
{\it Idle Time Assertion.}
Let $u$ be a node at height
$h_u\geq \log (c_2 B/c_1)$ in the up-tree.
The up-pass computation at $u$ is completed by time
$T_u = T + t + b  (c h -  k_u)$, where $h$ is the height of the up-tree,
and $k_u$ is the number of
writes {\it not yet performed} at proper ancestors of $u$ in the up-tree.

\vhalf

The proof is by induction on the height of $u$ in the up-pass tree, and
takes into account the fact that the entries
on any given execution stack is
a sequence of data on successive ancestors in the tree.

\vhalf
\noindent
{\it Base Case.} The node $u$ is at height
$\log (c_2 B/c_1)$.
Then, it is done
by time $T+t$.
Since $ch$ is the total number of writes that need to
be performed at all nodes on a path from a leaf to the root,
we have
$ch \geq k_w$
for every node $w$ in the up-tree, hence $T+t \leq T_u$, and hence $u$'s
computation is done by time $T_u$.

\vhalf
\noindent
{\it Induction Step.} Assume inductively that the result holds for all
nodes at height up to
$h-1 \geq \log (c_2 B/c_1)$.
Let $u$ be a node at height $h-1$,
let $v$ be $u$'s sibling, and let $w$ be $u$'s parent, which is at height
$h$.

By the inductive assumption, node $u$'s computation is completed
by
time
$T_u$. Further, since $u$ and $v$ have the same proper ancestors, $v$'s
computation is also completed by
time $T_u$, as is the computation at all
descendants of $u$ and $v$.

Now consider the computation at $w$. It is completed by time
$t_w = T_u + b \cdot (x + y')$, where $x$ is the number of accesses performed
by the task at $w$,
and $y'$ is the number of accesses performed
by tasks that write into
blocks shared by $w$.
By the inductive assumption all nodes at height $h-1$ and lower have
completed by time $T_u$, hence all of these writes are to locations at
proper ancestors of $w$.
Let $y$ be the
number of writes performed at all proper ancestors of $w$ during this time.
Then, $y' \leq y$, and $k_w = k_u - (x+y)$, hence,
\[
t_w = T+ t + b  (c h -  k_u) + b(x+y') \leq T +t + b (ch - (k_u - (x+y))) = T_w
\]

This establishes the induction step.

\vhalf
With this claim we have the result that the delay for the up-pass beyond
time $T$ is $O(b(B\log B + \log n))$, and this establishes the lemma.
\end{proof}

Let $s_P$ be the
delay at a core due to the
cost of a steal under PWS. When a computation with
critical pathlength $D$ is scheduled under PWS, it incurs a total
scheduling cost of $O(p \cdot s_P \cdot D)$ by Corollary \ref{cor4}.
We can expect $s_P \geq b$,
since at least one cache miss is incurred when a core attempts to steal
a task from another core.

\begin{lemma}\label{lem:idle}
Let $\pi$ be a BP computation on an input of size $n$.
If $s_P= \Omega (b)$,
then the total time spent on idle work by all $p$ cores in executing
$\pi$
is bounded by the sum of the bounds of $O(s_P \cdot p\log n)$ on
the scheduling cost and
$O(bBp \log B)$ on the block delay costs.
\end{lemma}
\begin{proof}
As shown above, the idle time for the up-pass is
$O(pb(B\log B + \log n))$.
The first term is dominated by the cost
$O(pB\log B)$ we established for
the PWS block miss excess and the
second term is dominated by the PWS scheduling cost if $s_P \geq b$.
The idle time that occurs in the top few levels of the downpass, until $p$
tasks are generated, is part of the steal cost $s_P$,
since the priority increases with each unsuccessful steal.
Hence the idle work is dominated by the sum of block
delay and scheduling costs.
\end{proof}

\subsection{Analysis of HBP Algorithms under PWS}\label{sec:app}

We apply our results for PWS
to the HBP algorithms discussed in Section \ref{sec:alg}.
Here, we assume that each core performs a single operation in $O(1)$ time,
and a cache miss takes at most $b$ time.
We assume that the input is of size
$n \geq Mp$ (the input size is $n^2$ for matrix computations), and it
is in the shared memory at the start of computation.

\vhalf
\noindent
{\bf Caches Misses Only.} If we ignore block misses, then for all
all Type 1 and Type 2 HBP algorithms
we consider,
PWS achieves the following run time, where
$s_c$ is the cost of a steal when only cache misses are
considered. We show that $s_C= b \log p$ for our PWS implementation
(Section \ref{sec:pws-impl})

\[
O\left( \frac{1}{p} \left(~ W(n) + b \cdot Q(n,M,B)~\right) + s_C \cdot T_{\infty}(n) \right) ~~\mbox{ if $M \geq B^2 $}.
\]

This is a new result, which is not known to be achievable using RWS.

\vhalf
\noindent
{\bf Cache and Block Misses.}
If we consider both cache and block misses, and $s_P$ is the cost
of a steal when both cache and block miss costs are considered, we
achieve the following bound with PWS for Type 1 and Type 2 HBP
algorithms we consider,
where $s_P = O(b \log p)$ for our PWS implementation if we use a {\it padded}
version of BP and HBP computations (see Section \ref{sec:pws-impl}), and
is $O(b (B + \log p))$ with standard HBP computations.

\[
O\left( \frac{1}{p} \left(~ W(n) + b \cdot Q(n,M,B)~\right) + s_P \cdot T_{\infty}(n) \right) ~~\mbox{ if $M \geq \Gamma(B)$}.
\]

Here the tall cache requirement $\Gamma(B)$ depends on the problem, and
varies between $B^2 \log B$ and $B^4$.

We now
present our results for our Type 1 and Type 2 HBP algorithms, when
scheduled under PWS.
For convenience of notation, we will ignore constant factors in our
analyses, and we will use $\geq$ and $\leq$ in place of $O$ and $\Omega$.

\begin{lemma}
When scheduled using PWS, the following algorithms have the stated
running times when the input size is $\Omega (Mp)$, considering both
cache and block misses.

(i) Scans: $O((1/p) \cdot (n + b \cdot (n/B) + s_P \cdot \log n)$ with
$\Gamma(B)= B^2 \log B$.

(ii) MT (BI) and RM to BI:
The bounds from (i) apply, with $n^2$ replacing $n$.

(iii) Strassen (BI):
$O((1/p) \cdot (n^{\lambda} + b \cdot (n^{\lambda}/(B \cdot  M^{\frac{\lambda}{2}-1})) + s_P \cdot \log^2 n)$
with $\Gamma (B) = B^2 \log^2 B $.

(iv) Depth-n-MM (BI):
$O((1/p) \cdot (n^3 + b \cdot (n^3/(B \sqrt M)) + s_P \cdot n)$
with $\Gamma(B) =  B^4$.

(v) BI-RM (gap RM):
$O((1/p) \cdot (n^2 + b \cdot (n^2/B) + s_P \cdot \log n)$
with $\Gamma(B)=  B^3 \log^2 B$.

(vi) BI-RM for FFT:
$O((1/p) \cdot (n^2 \log \log n + (b/B)  \cdot (n^2 \log_M n) + s_P \cdot \log n)$
with $\Gamma(B) =  B^2 \log B \log\log B$.

(vii) FFT:
$O((1/p) \cdot (n \log n + (b/B)  \cdot n \log_M n + s_P \cdot \log n \log \log n)$
with $\Gamma(B) =  B^2 \log B \log\log B$.

\vhalf
\noindent
All of these bounds hold with the standard tall cache $M \geq B^2$ if only
cache misses are considered.
\end{lemma}

\begin{proof}
$(i,~ii)$ {\bf Scans}, {\bf MT (BI)}, {\bf RM to BI}.
As noted in Section \ref{sec:alg},
we have
a single BP computation with $\alpha = 1/2$,
$f(r)= O(1)$, and $L(r)=O(1)$. Hence by Lemma \ref{lem:pws2},
the PWS cache miss excess is $O(Mp/B~+~p\log B)$,
and by Lemma \ref{lem:block},
the PWS block miss excess is $O(Bp \log B)$.
{\bf RM to BI} has $f(r) = \sqrt r$, however, Lemma \ref{lem:pws2} continues
to hold for all $f(r) = O(\sqrt r)$, hence this computation has the same
bounds as MT.
Hence the cache and block miss excess is
dominated by the sequential cache complexity
when $n \geq Mp$ if $M \geq \Gamma (B)$, for $\Gamma (B) = B^2 \log B$.

\vhalf
\noindent
$(iii)$ {\bf Strassen (bit-interleaved layout).}
From Section \ref{sec:alg} this is an HBP computation with
$c=1$, $r(m) = m/4$, where
$m=n^2$ is the size of the matrix, $f(r)=O(1)$ and $L(r) = O(1)$.
As noted there,
the sequential work is $O(n^{\lambda})$, the critical pathlength is
$O(\log^2 n)$,
and the sequential cache complexity is $O(n^{\lambda}/(B M^{\gamma}))$.
By  Lemma \ref{lem:c-miss-rec-perf},
substituting $s(n^2)=n^2/4$, and $c=1$,
the PWS cache miss excess is
$O(p[ \frac{M}{B} \log \frac{n^2}{M}  + \log^2 B ] )$,
and by Lemma \ref{lem:block-miss-rec-perf},
the PWS block miss excess is
$Q_B=O(pB\log B\log n^2)$.

The most constraining constraint is the block miss excess $Q_B$. We now derive
a bound for $\Gamma(B)$ based on $Q_B$.
We need $Q_B = pB \log B\log n^2 \leq \frac{n^{\lambda}}{BM^{\lambda/2 -1}}$,
which is satisfied if
$pB^2 \log B \cdot M^{\lambda/2 -1} \leq \frac{n^{\lambda}}{\log n^2}$;
thus $pB^2 \log B \cdot M^{\lambda/2 -1}\log(pBM) \leq n^{\lambda}$
suffices.

Since we have assumed that $n^2 \geq Mp$, we have
$n^{\lambda} \geq (Mp)^{\lambda/2}$.

Thus it suffices to have
$B^2 \log B (pM)^{\lambda/2} \frac{1}{p^{\lambda/2 -1}M} \log (pMB) \le (Mp)^{\lambda/2}$,
and this is met if
$\frac{B^2\log B}{M} \frac{\log p + \log M +\log B}{p^{\lambda/2 -1}} \le 1$.
As $\log M \ge \log B$ and $p^{\lambda/2 -1} \ge \log p$ (as $\lambda > 2.5$),
$\frac{\log p + \log M +\log B}{p^{\lambda/2 -1}}  \le \log M$.
Thus
$\frac{B^2 \log B \log M}{M} \le 1$ suffices, and for this the following stronger tall cache
condition $\Gamma(B) = B^2 \log^2 B \le M$ suffices.

\hide{
Since $\lambda > 2.5$, we have
$(Mp)^{\lambda/2 - \epsilon} \geq p \cdot M^{\lambda/2 -1} \cdot M^{1-\epsilon}
\geq p M^{\lambda/2 -1} B \log B$ if $M^{1-\epsilon} \geq B \log^2 B$
and
hence a standard tall cache $M \geq B^2$ suffices.
} 

\vone
\noindent
$(iv)$ {\bf Depth-n-MM (BI layout).}
Here
we use the cache-oblivious $n^3$-work MM algorithm in \cite{FLPR99},
as modified in \cite{CR11} to enforce limited access variables.
This is an HBP computation with $c=2$, and $r(n^2) = n^2/4$,
with $f(r)=O(1)$ and $L(r)=O(1)$,
hence by
part $(iii)$ of Lemma \ref{lem:c-miss-rec-perf}, the cache
miss excess is
$O(p(1 + \frac{n \sqrt{M}}{B}))$. The block
wait cost is $O(pnb B)$
by Lemma \ref{lem:block-miss-rec-perf}.
The cache miss excess and block delay cost are
bounded by the sequential cache complexity of
$\Theta (\frac{n^3}{B \sqrt M})$ for input size $n^2 \geq Mp$ with a
taller cache having $\Gamma (B ) \geq B^4$.

\vone
\noindent
$(v)$ {\bf BI-RM (gap RM).}
This is an $O(\log n )$ parallel running time, $O(n^2)$ operation
algorithm.

This is a sequence of two BP computations where the second computation has
the bounds of a standard scan, so the cost is dominated by the first
BP computation.
As described in Section \ref{sec:alg}, this first BP computation has
$f(r)=O(1)$, and the cache miss excess remains $O(Mp/B)$
for $M \geq B^2 \log B$.
For the block miss excess we have $L(r) = \sqrt r$. However,
its effect is reduced using a
gapping technique, the result of which is that there is no block miss
cost for tasks of size greater than
$\sigma = B^2 \log^4 B$.
The total block miss cost for stolen tasks of size $\sigma$
or less is dominated by the
stolen tasks of size $s$ due to the geometrically decreasing sizes in
the BP computation. Hence the block miss excess is $O(p \cdot B^2 \log^2 B)$.
Since the sequential cache complexity is $n^2/B$, this dominates the block
miss excess for input size $n^2 \ge  pM$ with a taller cache
$\Gamma(B) \geq B^3\log^2 B$.

\vone
\noindent
$(vi)$ {\bf BI-RM for FFT.} $O(\log n )$ parallel running time,
$O(n^2\log\log n)$ operation algorithm.

Recall from Section \ref{sec:alg} that this
is a type 2 HBP
computation with $c=1$, with $s(n^2) = n$.
From Section \ref{sec:alg} we have $f(r) = O(\sqrt r)$ and $L(r) = O(1)$.
This gives a cache miss excess of
 $O(p\frac{M}{B} \log\log_M {n^2}  )$.
By lemma~\ref{lem:block-miss-rec-perf},
the excess block wait cost is
$O(pB \log B \log \log B)$.
This is dominated by the sequential cache complexity of FFT (on an input
of size $n^2$) if $(n^2/B) \log_M n^2 \geq pB \log B \log\log B$, and
this is satisfied if
$n^2 \geq pB^2 \log B \log\log B$. Hence a tall cache
$M \geq B^2 \log B \log\log B$ suffices.

We will see below that
as the costs of this method are dominated by those for FFT in all dimensions
(work, parallel time, and cache and block misses) this algorithm can be used
for FFT.

\vone
\noindent
$(vii)$ {\bf FFT.}
This is a Type 2 HBP computation with $c=2$ and $s(n) = \sqrt n$.
We have
$O(1)$-friendly tasks with $L(r)=O(1)$
(outside of the cost to finally convert the BI matrix into the RM output
format).
By Lemma \ref{lem:block-miss-rec-perf},
the excess block wait cost is
$O(pB\log n \log \log B)$.
At the end, to convert to the RM format
we use {\bf BI-RM for FFT)}.
\end{proof}


\subsection{List Ranking}\label{sec:aapp2}

We now analyze the cache miss, block miss and idle time cost overheads
for the list ranking algorithm LR presented in Section \ref{sec:alg}
when scheduled by PWS. For this, we
need the following extension to the results for SPMS in \cite{CR10b}.

\begin{lemma}
\label{lem:small-sort-costs}
The excess cache miss costs for sorting $x$ items, $p < x <pM$,
using SPMS, when scheduled under PWS
is bounded by
\[
O\left( \frac{x}{B} \log \frac{pM}{x} \frac{\log x}{\log x/p} + \sqrt{xp}\frac{\log x}{\log x/p} \right).
\]
\end{lemma}
\begin{proof}
The excess cache miss cost of a BP collection of size $r$ and total size $x$ is given by
\begin{eqnarray*}
O(\left( \frac{y}{B} \log \frac{pM}{y} + \sqrt{yp} + p\log y/p \right) & ~~~~~r\ge y/p \\
O(p\sqrt{r} + p\log r) & ~~~~~r < y/p
\end{eqnarray*}
\hide{
The excess cache miss cost of a BP collection of size $r$ and total size $x$ is given by
\begin{eqnarray*}
O(\left( \frac{x}{B} \log \frac{pM}{x} + \sqrt{xp} + p\log x/p \right) & ~~~~~r\ge x/p \\
O(p\sqrt{r} + p\log r) & ~~~~~r < x/p
\end{eqnarray*}
} 
For the sorting problem, there are some c=O(1) collections of
size $x$, $2c$ of
size $\sqrt{x}$, and in general,
$2^i c$ of
size $x^{1/2^i}$, for $1 \le i \le \log\log x$.
All these collections have total size $O(x)$.
Summing over all values of $i$ yields an excess cache miss cost of
\[
O \left( \frac{x}{B} \log \frac{pM}{x} \frac{\log x}{\log x/p}
  + \sqrt{xp}\frac{\log x}{\log x/p} + p \log x \log\log x/p \right).
\]
The result follows, because the second term dominates the third one.
\end{proof}

\begin{corollary}
\label{cor:cache-miss-lr}
The cache miss cost in the list ranking algorithm is
$O(\frac{n}{B} \frac{\log n}{\log M} +
 \frac{Mp}{B} \frac{\log n}{\log M} \log^{(k)} n )$,
for
inputs of size $n\ge Mp \log^{(k)} n$, assuming that
$M\ge B^2\log B \log\log B$.
\end{corollary}
\begin{proof}
In each iteration of the list ranking algorithm there are $O(\log^{(k)}n)$
invocations of the sorting algorithm.
As shown in \cite{CSBR09},
the $O(Q)$ terms over all invocations of the
sort procedure sum to $O(\frac{n}{B} \cdot  \frac{\log n}{\log M})$.
Thus it suffices to add the costs given by Lemma
\ref{lem:small-sort-costs}
for halving values of $x$, starting at $x=n$, and then multiply the
resulting sum
by $\log^{(k)} n$.

We consider this sum in more detail.
For $x\le p$, these terms contribute a total of
$O(p\log p) =O(\frac{Mp}{B}\frac{\log n}{\log M} \frac{B\log M}{M})$
$=O(\frac{Mp}{B} \frac{\log n}{\log M})$.

 For $p\le x \le Mp$,
The terms $O(\sqrt{px} \frac{\log x}{\log (x/p)})$ contribute
$O(\sqrt{pMp}  \frac{\log Mp}{\log M})=O(p\sqrt{M} \frac{\log n}{\log M})=O(\frac{pM}{B}\frac{\log n}{\log M})$, with a tall cache $M \geq B^2$.
The terms $O(\frac{x}{B} \cdot \log (Mp/x) \cdot \frac{\log x}{\log (x/p)})$
contribute
$O(\frac{Mp}{B} (\frac{\log Mp}{\log M} + \frac{\log 2}{2} \frac{\log (Mp/2)}{\log M/2}
                            + \frac{\log 4}{4} \frac{\log Mp/4}{\log M/4}
                            + \cdots + \frac{\log M}{M} \frac{\log p}{1} )$
   $=O(\frac{Mp}{B} \frac{\log n}{\log M})$.
\end{proof}

For the block wait cost, we note that with the gapping method,
no more block misses occur once the list has size at most $n/B^2$.
Recalling that the sort algorithm has a block miss cost of
$O(Bp\log n \log\log (n/p) )$ to
sort a set of $n$ items \cite{CR10b},
we obtain the following block delay cost for the list ranking algorithm.

\begin{lemma}
\label{lem:lr-block-miss}
The list ranking algorithm has a block wait cost of
$O(Bp\log n\log\log (n/p) \log^{(k)}n\log B)$.
\end{lemma}
\begin{proof}
This bound follows from observing that there are
$O(\log^{(k)}n)$ sorts on
sets of size at most $n$ in each iteration of the list ranking algorithm.
Further, only the first $O(\log B)$ iterations incur block misses.
\end{proof}

\begin{lemma}
\label{lem:lr-seq-comp-constraints}
The cache and block wait
costs of the list ranking algorithm are bounded
by the sequential cache miss cost of $O(\frac{n}{B}\frac{\log n}{\log M})$
if $Mp\le n/\log^{(k-1)} n$ and $B^2 \log^2 B \log\log B
\le M$.
\end{lemma}
\begin{proof}
The claim regarding the cache miss cost is immediate from Corollary
\ref{cor:cache-miss-lr}.
To show the claim for the block miss cost,
by Lemma \ref{lem:lr-block-miss}, it suffices to show that
$O(Bp\log n\log\log (n/p) \log^{(k)}n\log B) \le
O(\frac{n}{B}\frac{\log n}{\log M})$.
That is, $O(B^2 \log M \log^{(k)}n\log B) \le
O(\frac{n/p}{\log\log n/p})$.
Then it suffices to show that
$O(B^2 \log M \log^{(k)}n\log B (\log\log B + \log^{(3)}M + \log^{(k+2)} n ) \le
O(n/p)$.
If $Mp \le n/\log^{(k-1)} n$, then it suffices that
$O(B^2 \log M \log B (\log\log B + \log^{(3)} M) ) \le M$,
for which $B^2 \log^2 B \log\log B = O(M)$
suffices.
\end{proof}

Finally, we obtain
the following result for the list ranking algorithm when
scheduled under PWS.

\begin{theorem}\label{thm:lr}
If $M \geq B^2 \log^2 B \log\log B$ and $n\geq Mp \cdot \log^{(k-1)} n$, then
the list ranking algorithm runs in time
\[
O\left( \frac{1}{p} \left(~ n \log n + b \cdot sort(n)\right) + s_P \cdot \log^2 n \log\log n \right)
\]
\end{theorem}
\begin{proof}
We only  need to bound the idle time. For this, we note that the idle time
for computation on each contracted list of
size greater than $p$ is bounded by the corresponding bounds
for sorting and for scans. When the contracted list has size $x<p$, the
computation proceeds with full parallelism, and the idle time is
$O((b \cdot p \log x \log\log x)$, for a total cost of
$O((b \cdot p \log^2 p \log\log p )$ for computation on all
contracted lists of size at most $p$. Since $p<n$, this is bounded
by the inherent
cost of each parallel step in the computation, and this establishes
the bound in the theorem since it is
obtained by the bounds derived earlier for the work, critical pathlength,
cache miss cost and block miss cost.
\end{proof}

The Euler tour and tree computation algorithms have
the same complexity since they are simple applications of the parallel
list ranking algorithm.

Finally, the dominant cost in the
connected components algorithm in \cite{CSBR09} is
$\log n$ stages of list ranking, and we obtain resource-oblivious
implementation under PWS with both parallel time and cache complexity
increased by a factor of $\log n$.

\subsection{Distributed PWS implementation}
\label{sec:pws-impl}

We present a simple distributed implementation of PWS.
This implementation assumes that each core context switches from its
actual computation to perform $O(1)$ computation
on the distributed implementation every $k$ time units, for
a suitable $k$.
If $k$ is a constant, then
each core will devote a constant fraction of its run time to
tending to scheduling issues.

Our implementation
supports steals
by maintaining two full binary trees
on $p$ leaves,
the {\it steal tree} $S$ and the {\it task tree} $T$,
with the $i$th leaf in each tree for the $i$th core. The
steal tree and task tree
support a prefix sums BP computation on a
steal array $S[1..p]$
and task array $T[1..p]$ respectively.
There is a pointer from
$S[i]$ to a location that
is set to 1 if process $i$ needs to steal,
and similarly there is a pointer from $T[i]$ to
the task at the head of core $i$'s task queue that is available to be
stolen.

The scheduling
proceeds in phases, each taking $O(\log p)$
steps, and a steal request
by a core will be processed in the phase that starts after the current
phase completes. The $i$th core is responsible for the computation at
leaf $i$, and at most one pre-assigned internal node in each tree.
A scheduling phase starts in tree $S$, where each {\it active}
core $i$ (i.e, core that needs to
steal) assigns a 1 to $S[i]$. The remaining positions are assigned 0.
The $p$ cores compute prefix sums in $O(\log p)$
steps, after which each active
core knows its rank among the active cores.
Then, the computation proceeds to the task tree.
Those tasks in $T[i]$ with priority matching the priority of the
current round are assigned a 1, and the
remaining leaves are assigned 0. A prefix sums computation on this tree
assigns ranks to the tasks available to be stolen. The steals and tasks
then match up by rank and the steals proceed to be executed, while
the scheduling moves to the next phase.

One additional case that can occur in the task tree computation
arises when a non-idle
core $i$ needs to assign the priority of task $T[i]$ in the task
tree, but the task queue of core $i$ is empty, and the core has not yet
generated its next forked task, if any.
In this case core $i$ assigns the priority of
the node it is currently executing minus 1
as the priority of the task for $T[i]$,
and it sets a flag to indicate that this value is
an upper bound on the
priority of a task that has not yet been generated on the task queue.
If this priority matches the priority of the current round, then the
stealing cores wait till either
a task is generated on core $i$'s task queue
or core $i$ becomes idle;
otherwise this flagged task is ignored since its priority is lower than
that for the current phase.

\subsubsection{Analysis}

Since each step may entail a non-local read, the time for a step
is at least the time for $\Theta(1)$ cache misses.
The only block miss costs are those that occur in accessing the arrays
storing $S$ and $T$.
At the beginning of the computation,
each core requests space to store the entries at the leaves and internal
nodes in $S$ and $T$ that it accesses. This is a total of at most four items
to a core, and a different block is assigned to each core by our space
allocation property. Hence
accessing any item incurs $O(1)$ block wait cost.
Thus, the delay to a stealing core is $O(b\log p)$ within this
scheduling computation.

\vhalf
\noindent
{\bf Block Wait Time at Execution Stacks.}
As mentioned above, there could be
an additional delay associated
with a steal, and this is the block wait time that could be
incurred by the stealing core while it waits for a core to place a task
on a task queue when that queue is empty. Our analysis on block wait costs
in an HBP computation bounds this delay to $O(B)$, thus resulting in the total
delay due to block waits
in an HBP computation with critical pathlength $T_{\infty}$
being
$O(pB\cdot T_{\infty})$. Hence the overall cost of all steals is
$O(pb (B + \log p) T_{\infty})$.

\vhalf
\noindent
{\bf Reducing the Block Wait Cost of Steals.}
We describe here a method to reduce the
overhead of the cost of steals in HBP computations by using
padded BP and HBP computations, which were defined in
Definitions \ref{def:paddedBP} and \ref{def:rec-alg}.

All of our analyses for cache and block misses continue to hold
for padded BP and HBP computations, since each empty array is placed
between two different segments on an execution stack, and this can only
reduce the access costs.
Additionally, block wait costs are reduced to
$O(1)$ at nodes of
height
$\Omega(\log B)$ in any BP computation, since
when
the allocated empty arrays are of length greater than $B$, there is
no interaction between accesses to segments for different nodes on the stack.
As a result, the total block wait of all steals under PWS in a
padded BP computation of size $r$ reduces to
$O(p b (\min\{r\log r, ~B \log B\} )$. This cost is absorbed in
the block miss cost of the actual execution of the BP computation as shown
in Lemmas \ref{lem:block} and \ref{lem:up-block}.
Similarly, by Lemma \ref{lem:block-miss-rec-perf}, the block wait cost
of steals under PWS in Type 2 padded HBP computations for all algorithms
we consider in this paper is dominated by the block wait cost incurred by
the actual computations.

Thus, if we use padded BP and HBP computations for our algorithms then
the overhead for the scheduling and the steals under our PWS distributed
implementation is $O(pb\log p \cdot T_{\infty})$.
Hence, the excess in the cost of cache and block misses in this PWS
implementation over the cost of the
cache and block misses in the HBP computation is the same as its
cache miss excess when considering only
the cost of cache misses in the HBP computation.

We note that if we use padded BP and HBP computations in place of
standard BP and HBP computation, the block miss costs decrease even in
the actual computations. However, the reduction
in block misses in the downpass of each BP computation occurs only at
higher levels of the BP tree, where the inherent cache miss cost of the
computation is high, as shown in our analysis in Section \ref{sec:block-pws}.
In the up-pass, again as shown in Section \ref{sec:block-pws}, the total
block miss cost in a standard BP computation is dominated by the block miss
cost of the downpass, at least for $\alpha=1/2$, and a padded BP
computation offers no benefit beyond a constant factor.
Padded computations may come in useful in
other HBP algorithms, for instance those that use $\alpha > 1/2$.

\vhalf

Other simple implementations are possible
for PWS, and may be more practical when the input size is much larger
than the number of cores.
For instance, by using a `prisoner type' computation \cite{FRW88},
each scheduling round can complete in $O(\log p)$
steps, with only the
active cores involved in the computation (in contrast to the first method
where each core would need to devote a constant fraction of its time to
the scheduling process).
Here an active core is one that needs to steal or one for which the
task at the head of its task queue has changed.
The available tasks will be grouped in sets, one
for each priority.
This
second method would be more efficient
as long as variation in processor delays is not too large, since each
core would need to wait the maximum time at each level of the tree where
its sibling is not active, in order to ensure that no value is being
propagated up the tree by the sibling.

\vhalf
There is one more point to note: if computation by the cores can continue
during the execution of steal requests, a task available at the start of a
phase may no longer be available when it is assigned to be stolen.
We observe that this does not affect our algorithmic
analysis. Let us call such a task a \emph{pseudo-stolen} task.
Then, for each priority $d$, there will be at most $p-1$ tasks of
priority $d$ that are stolen or pseudo-stolen. The remainder of the
analysis now proceeds as before.

\section{Discussion}\label{sec:disc}

\subsection{Other Mechanisms to Handle Block Interference}
\label{sec:blck-disc}

For many of our algorithms, the
tall cache requirement that $\Gamma (B)$ be in excess of $B^2$
is imposed by the block misses. Here we
consider some strategies that could be utilized by the system hardware
or software that could potentially mitigate this cost.

\vone
1. {\it 2-Core Block Sharing.}
It would be helpful if the operating
system could help prevent ping-ponging by allowing one core
to do all its writes before the other core accesses a pairwise
shared block.
The second core would then face a more acceptable delay of $O(B)$
operations and one cache miss. This might be implemented by a lock
with a delayed release, though one would need to be careful to avoid
deadlock. However, this is beyond the scope of the present paper.

2. {\it Many-Core Block Sharing.}
This occurs when there are tasks with very small
memory footprint. Then, inevitably, multiple tasks will be
writing into the same block. This seems to be unavoidable in an
oblivious setting, for the algorithm designer cannot set a
minimum task size in terms of the block size.\footnote{Of course,
the algorithms designer may create a minimum size to balance the
operation overhead of task creation; these would be costs found even
in a uniprocessor execution of that algorithm and even without considering
a cache-oblivious execution.}

In fact,
this issue seems to be even more salient in the context
of a hierarchically organized collection of cores
(see below),
an organization
that seems inevitable as the number of cores grows. For that scenario,
a mechanism was proposed in \cite{CSBR09}
whereby the algorithm provides the run time scheduler a space bound
for each task, and using this, a method for scheduling tasks
effectively on a multi-level memory hierarchy was given.
  (see also
 the discussion in Section \ref{sec:hierarchy}).
A similar
mechanism could be used to avoid stealing unduly small tasks,
thereby eliminating the multi-way block sharing costs for the most part.

\subsection{Hierarchy of Caches}\label{sec:hierarchy}

We have so far assumed that the caching environment in the multicore
consists of a private cache for each core, all of which access data from an
infinite shared memory. As the number of cores $p$ gets large
it is to be expected that a caching hierarchy of $d$ levels would
be present, where each cache at level $i$ is shared
by some number of caches at level $i-1$, $d \geq i \geq 2$.
In actual practice $d=2$ is a common configuration currently,
where each core has a private cache ($L_1$) of size
$M_1$ and all $p$ caches share a shared cache $L_2$
of size $M_2 > p \cdot M_1$, and this shared cache accesses data from the
infinite global shared memory.

A simple (but non-optimal way) of utilizing a level-$i$ cache $Q_i$ that is
shared by $k$ level-$(i-1)$ caches is to consider $Q_i$ as being
partitioned into $k$ disjoint segments of equal size, and assigning
each segment to one of the level-$(i-1)$ caches.
If the properties assumed by the sequential cache oblivious analysis hold
at every cache in this hierarchy, then our algorithms
can guarantee optimal use of this partitioned
cache hierarchy.
This follows from the results for sequential
cache-oblivious algorithms, since each private cache can be viewed
as having its own cache hierarchy consisting of a proportionate
portion of each cache it shares at each level of the hierarchy.
This observation is also noted in \cite{BGS09}.
However, one could hope to do much better by having
all cores that share a given level-$i$ cache compute
on different parts of the same subproblem so that the data is
utilized more effectively. Given the competing demand for
private versus shared caches, this appears to be
a challenging task in a completely resource-oblivious
environment. It appears that a mechanism, such
as that proposed in \cite{CSBR09}, is needed. The
approach there is to have a resource-oblivious
specification of the algorithm along with some basic
`hints' to the run-time scheduler, which allows for
optimal utilization of shared caches at all levels of
the cache hierarchy by a scheduler that is aware of the
cache sizes.

\subsection{HBP Algorithms in a Bulk-Synchronous Environment}

A bulk-synchronous environment is used in \cite{AGN08, Va08} to
develop efficient, parameter-aware multicore algorithms. We observe
here that
balanced HBP algorithms
can be mapped efficiently
on to these models, as described below.

If the computation is a pure BP
computation of size $n$, then we fork $\log p$ levels of recursion to
obtain $p$ tasks of size $n/p$ which are mapped on to the $p$ cores.
There is a synchronization after the $p$ cores complete this part of
the computation,
followed by $\log p$ supersteps to complete the computation for the
top $\log p$ levels. (This can be further refined to obtain a better
bound as a function of $L$ and $g$, using the usual BSP methodology,
if these parameters are used in the model).

For an HBP computation, we
unravel
recursive calls until the first level of
recursion where the size of the subproblem becomes smaller than the
cache size, and at least $p$ parallel recursive tasks have been
generated,
where $p$ is the number of cores if only private caches are considered,
and is the number of parallel caches at this level of the cache
hierarchy, otherwise.
We then complete the entire computation in terms of
subproblems of this size, each of which is computed cache efficiently
on the cores. The larger BP computations within this sequence of
computations are handled as described in the above paragraph. On a
multi-BSP we perform this process of determining recursion level
at each
level of the cache hierarchy, starting with the top level. On a
multicore with a private cache, executing bulk-synchronously,
this computation can be made
cache-oblivious by unfolding the recursion to the first level where
the total number of subproblems become $p$ or greater (rather than
using the cache size to determine the level of recursion).

We note
that multicore-oblivious versions of many of the HBP algorithms we
have presented are
given in \cite{CSBR09} for multi-level cache hierarchies, and some
of these
algorithms are quite similar to the corresponding algorithms
in \cite{Va08}; all of these algorithms operate along the lines of
the method we have described above.

\section{Conclusion}

Our results are essentially optimal, subject to the algorithm with which
we are starting, except for the
modest overhead for the steals (the cost for $s_P$).

We have observed that many well-known algorithms are intrinsically HBP
algorithms (and hence limited access). The one exception is Depth-n-MM,
where we used the modified version given in \cite{CR11} instead of the
algorithm in \cite{FLPR99} because the latter, being in-place, has $n$
writes to each of the $n^2$ output locations, and hence is not limited
access. The modified algorithm in \cite{CR11}
has the same work and cache complexity,
but achieves limited access by using local variables, and BP computations
for copying. The same approach can be used to obtain limited access
versions of I-GEP and LCS \cite{CR06}. In all cases, we retain the work
and cache bounds, while losing
the in-place property, though the additional space used is for local variables, and this
space is re-used during the computation.

\bibliographystyle{abbrv}

\bibliography{sort,refs}

\end{document}